\documentclass[11pt,document,nofootinbib,superscriptaddress,onecolumn,preprintnumbers,balancelastpage]{article}

\pdfoutput=1
\usepackage{jheppub}
\usepackage{graphicx}
\usepackage{epstopdf}
\usepackage{dcolumn}
\usepackage{bm}
\usepackage{hyperref}
\usepackage{booktabs}
\usepackage{enumitem}
\usepackage{blindtext}
\usepackage[dvipsnames]{xcolor}
\usepackage{amsmath}
\usepackage{cancel}
\usepackage{xpatch}
\usepackage{mathtools}
\usepackage{amsthm}
\definecolor{c1}{rgb}{0.368417, 0.506779, 0.709798}
\definecolor{c2}{rgb}{0.880722, 0.611041, 0.142051}
\definecolor{c3}{rgb}{0.560181, 0.691569, 0.194885}
\definecolor{c4}{rgb}{0.922526, 0.385626, 0.209179}
\definecolor{c5}{rgb}{0.528488, 0.470624, 0.701351}
\definecolor{c6}{rgb}{0.772079, 0.431554, 0.102387}
\definecolor{c7}{rgb}{0.363898, 0.618501, 0.782349}
\definecolor{turq}{rgb}{0.181,0.638,0.594}
\definecolor{pink}{rgb}{1.000,0.54,0.8}

\def\beq{\begin{align}}
\def\eeq{\end{align}}
\newcommand{\gsim}{ \mathop{}_{\textstyle \sim}^{\textstyle >} }
\newcommand{\lsim}{ \mathop{}_{\textstyle \sim}^{\textstyle <} }
\newcommand{\vev}[1]{ \left\langle {#1} \right\rangle }

\newcommand{\EV}{ {\rm eV} }
\newcommand{\KEV}{ {\rm keV} }
\newcommand{\MEV}{ {\rm MeV} }
\newcommand{\GEV}{ {\rm GeV} }
\newcommand{\TEV}{ {\rm TeV} }

\def\mpl{M_{\rm Pl}}

\def\GeV{{\rm GeV}}

\title{Sterile Neutrino Dark Matter in Left-Right Theories}

\author[1,2]{Jeff A. Dror}
\author[1,2]{David Dunsky}
\author[1,2]{Lawrence J. Hall}
\author[3]{Keisuke Harigaya}
\affiliation[1]{Department of Physics, University of California, Berkeley, California 94720, USA}
\affiliation[2]{Theoretical Physics Group, Lawrence Berkeley National Laboratory, Berkeley, California 94720, USA}
\affiliation[3]{School of Natural Sciences, Institute for Advanced Study, Princeton, New Jersey 08540, USA}

\abstract{
$SU(2)_L \times SU(2)_R$ gauge symmetry requires three right-handed neutrinos ($ N _i $), one of which, $N_1$, can be sufficiently stable to be dark matter. In the early universe, $ W _R $ exchange with the Standard Model thermal bath keeps the right-handed neutrinos in thermal equilibrium at high temperatures. $N_1$ can make up all of dark matter if they freeze-out while relativistic and are mildly diluted by subsequent decays of a long-lived and heavier right-handed neutrino, $N_2$. We systematically study this parameter space, constraining the symmetry breaking scale of $SU(2)_R$ and the mass of $N_1$ to a triangle in the $(v_R,M_1)$ plane, with $v_R = (10^6 - 3 \times 10^{12})$ GeV and $M_1 = (2\, \KEV - 1 \, \MEV)$. Much of this triangle can be probed by signals of warm dark matter, especially if leptogenesis from $N_2$ decay yields the observed baryon asymmetry. The minimal value of $v_R$ is increased to $10^8 \, \GEV$ for doublet breaking of $SU(2)_R$, and further to $10^9 \, \GEV$ if leptogenesis occurs via $N_2$ decay, while the upper bound on $M_1$ is reduced to 100 keV.  In addition, there is a component of hot $N_1$ dark matter resulting from the late decay of $N_2 \rightarrow N_1 \ell^+ \ell^-$ that can be probed by future cosmic microwave background observations. Interestingly, the range of $v_R$ allows both precision gauge coupling unification and the Higgs Parity understanding of the vanishing of the Standard Model Higgs quartic at scale $v_R$. Finally, we study freeze-in production of $N_1$ dark matter via the $W_R$ interaction, which allows a much wider range of $(v_R,M_1)$.
}

\begin{document} 
\maketitle
\flushbottom

\tableofcontents

\section{Introduction}

Left-right (LR) symmetry~\cite{Pati:1974yy,Mohapatra:1974gc,Senjanovic:1975rk} is a possible remnant of grand unification~\cite{Georgi:1974my,Fritzsch:1974nn,Georgi:1979ga,Kibble:1982dd}, can restore space-time parity at high energies, solve the strong CP problem~\cite{Beg:1978mt,Mohapatra:1978fy,Babu:1988mw,Babu:1989rb,Hall:2018let}, and explain the small Standard Model (SM) Higgs quartic coupling at high energy scales~\cite{Hall:2018let,Dunsky:2019api,Hall:2019qwx,Dunsky:2019upk}. In LR theories, the electroweak gauge group, $SU(2)_L \times U(1)_Y $, is extended to $SU(2)_L \times SU(2)_R \times U(1)_{  B - L }$, which is broken at a scale ($ v _R $) above the weak scale ($ v  $), $ v _R \gg v  $. LR symmetry predicts right-handed neutrinos, which, if their masses and mixing with the left-handed neutrinos are sufficiently small, can be stable. Since right-handed neutrinos are inert under the SM gauge group, they are candidates to make up the observed dark matter (DM) density of the universe. Right-handed neutrino DM belongs to a class of sterile neutrino DM, and we use ``right-handed neutrino" and ``sterile neutrino" interchangeably.

How can right-handed neutrino DM be populated in the early universe? For large $v_R$ and/or small reheating temperatures of the universe, production of right-handed neutrinos through the exchange of heavy gauge bosons, $W_R$ and $Z_R$, is negligible. Right-handed neutrinos can still be produced by their Yukawa coupling with the SM lepton douplets and Higgs~\cite{Dodelson:1993je}. However, this production mechanism is in tension with the constraints from x-ray searches and structure formation of the universe~(see e.g.~\cite{Perez:2016tcq}), unless a significant lepton asymmetry is present~\cite{Shi:1998km}.

Production of right-handed neutrinos by the exchange of $W_R$ and $Z_R$ becomes increasingly effective for higher reheating temperatures. The resultant abundance reproduces the observed DM density for an appropriate reheating temperature; above this temperature, right-handed neutrinos are overproduced.

In the limit of high reheating temperatures, right-handed neutrinos are thermalized via $W_R$ and $Z_R$ exchange.
The DM phenomenology of LR theories in the case of high reheat temperatures was first studied in~\cite{Bezrukov:2009th}, which showed that the lightest right-handed neutrino can make up DM if it decouples while relativistic and has its abundance diluted by decays of heavier right-handed neutrinos into the SM bath through an off-shell $W_L$ via sterile-active mixing.
The requirement that the heavier neutrino freezes-out while relativistic leads to a constraint on the $ W _R $ mass, $ M _{ W _R } \gtrsim 10^4  ~{\rm GeV} $,
with no clear upper bound.

In this work we study the parameter space of LR models systematically, mainly for reheat temperatures after inflation above the temperatures needed to thermalize the right-handed neutrinos by $W_R$ and $Z_R$ exchange. As in~\cite{Bezrukov:2009th}, right-handed neutrinos decouple relativistically, and the unstable but long-lived states decay to dilute the abundance of the stable state to the observed DM abundance. We extend previous work, finding a bounded parameter space from a combination of constraints including enough dilution, Big-Bang Nucleosynthesis (BBN), warm DM, hot DM, and $ \Delta N _{ {\rm eff}} $. Upper bounds on the DM neutrino mass and on the $SU(2)_R$ symmetry breaking scale, $v_R$, result from a detailed analysis of the neutrino mass matrix, which takes a form constrained by LR symmetry. Furthermore, the mass of the lightest active neutrino is constrained to be $\lesssim 10^{-4}$ eV. We discuss how the resulting parameter space will be probed observationally, especially using 21 cm cosmology, and also how it is further constrained if decays of the long-lived right-handed neutrino generate the observed baryon asymmetry via leptogenesis.  The range of $v_R$ predicted by the DM abundance is compared to ranges which lead to precision gauge coupling unification and to the observed value of the Higgs boson mass.

In addition, we study the case of lower reheating scales, finding that freeze-in is also a viable option to produce relic right-handed neutrinos. In this case, the sensitivity to the reheat temperature after inflation leads to a wide open parameter space, with values of $v_R$ as large as the Planck scale.

\section{Left-right models and neutrino masses}
\label{sec:nuinLR}
In this section we summarize the neutrino sector of left-right theories, emphasizing the role played by the LR symmetry. We begin by considering the effective theory of the SM with $ 3 $ additional gauge singlets, $ N _i $, and then introduce LR symmetry. The leading operators in the SM that give rise to masses for neutrinos are bilinear in lepton fields,
\begin{align}
-{\cal L}_{ \rm SM +N, \, eff}\; \supset   \; y_{ij} \,(\ell_i \,  N_j) \, H_L  +  \frac{y'_{ij}}{\Lambda} \; (\ell_i \, \ell_j )\, H_L^2  + y''_{ij} \, M_R \, ( N_i  \, N_j)    +  {\rm h.c.} \,. 
\label{eq:SMEFT}
\end{align}
where $\ell_i \equiv ( \nu_i , e_i ) $ are the three lepton $SU(2)_L$-doublet fields.  This involves three independent dimensionless flavor matrices $(y,y',y'')$ and two mass scales: the SM cutoff scale $\Lambda$ and the right-handed neutrino mass scale, $M_R$.   Without $N_i$, the SM only contains the second of these three operators~\cite{Weinberg:1979sa}, which is sufficient to adequately describe the observed neutrino masses and mixings, once the SM Higgs field $H_L$ acquires its vacuum expection value, $v$. When including $N_i$, the second term of (\ref{eq:SMEFT}) is often neglected, resulting in light neutrino masses from the seesaw mechanism~\cite{Yanagida:1979as,GellMann:1980vs,Minkowski:1977sc,Mohapatra:1979ia} if $M_R \gg v$.

In this paper we study the extension of the SM electroweak gauge group to $SU(2)_L \times SU(2)_R \times U(1)_{B-L}$. This simplifies the representation structure of the quarks and leptons: $q \equiv (u,d)$ and ${\ell} \equiv ( \nu , e )$ transforming as (2,1) under $SU(2)_L \times SU(2)_R$ and  $\bar{q} \equiv ( \bar{u} , \bar{d} )$ and $\bar{\ell} \equiv (N , \bar{e})$ transforming as (1,2). The presence of the right-handed neutrinos is now required by the gauge symmetry, and this is their natural setting. We impose a discrete symmetry that interchanges $SU(2)_L \leftrightarrow SU(2)_R$; the corresponding transformation on the fermions may include spacetime parity, $\ell \leftrightarrow \bar{\ell}^\dagger$, or not, $\ell \leftrightarrow \bar{\ell}$.

We do not specify the full structure of the LR symmetric theory, though any such theory must have the $SU(2)_R \times U(1)_{B-L}$ gauge symmetry broken to hypercharge at some scale $v_R \gg v$. We consider the effective field theory below $v_R$, assuming that the only fermions relevant for neutrino masses in the effective theory are $\nu_i$ and $N_i$, and the lepton-number violating contribution to their masses is generated by a single type of LR symmetric interaction. In this case, the leading operators for neutrino masses are
\begin{align}
-{\cal L} _{ \rm LR, \, eff}\; \supset   \;  y_{ij} \,(\ell_i \,  N_j) \, H_L  +   \frac{c \, y'_{ij}}{v_R} \, (\ell_i  \, \ell_j) \, H_L^2  +  y_{ij}'^{(*)} \,v_R \,  (N_i  \, N_j)    
+  {\rm h.c.} \,. 
\label{eq:LREFT}
\end{align}
If the LR symmetry includes spacetime parity, $y$ is a Hermitian matrix and the complex conjugation is included in the last term; otherwise $y$ is symmetric and the complex conjugation is omitted. Even though LR symmetry has been spontaneously broken, the $(\ell_i  \, \ell_j)$ and $ ( N_i N_j ) $ flavor matrices are identical, $y''_{ij} = y'_{ij}$, reflecting the symmetry structure of the full theory. This will have important consequences for the parameter space in which $N_1$ can be DM.   Furthermore, comparing with \eqref{eq:SMEFT} we find that $M_R= v_R$ and $\Lambda = v_R/c$, where the constant $c$ is discussed below, and is unity in certain theories. 

The effective Lagrangian leads to a $6 \times 6$ neutrino mass matrix,
\begin{align}
\begin{array}{c} \big( \begin{array}{cc}\nu  _i   & N _i \end{array} \big) \\ {} \end{array}
  \begin{pmatrix}
c M_{ij} \,v ^2 /v_R ^2  & y_{ij} v \\
 y_{ji} v & M_{ij}^{(*)}
\end{pmatrix} \bigg( \begin{array}{c} 
 \nu _j \\  
 N _j 
\end{array} \bigg)
  \,,
\label{eq:numassmatrix0}
\end{align}
where $M_{ij} = y'_{ij} v_R$.  Without loss of generality we can work in a basis where $y'$ is diagonal such that,
\begin{align}
	M_{ij} &= M_i \, \delta_{ij},
\end{align}
with all $M_i$ real.  Upon integrating out the three heavy states, we obtain a mass matrix for the three light neutrinos:
\begin{align}
	m_{ij} \, &= \, \delta_{ij}c \frac{v^2}{v_R^2} M_i - y_{ik} v \; \frac{1}{M_k} \; y_{jk} v \,\equiv \, \delta_{ij} \, m_{\nu,i}^{(5)} - m_{\nu,ij}^{(ss,N)}.
	\label{eq:numassmatrix}
\end{align}
In this basis, in the limit that $y _{ij}$ is diagonal the lepton flavor mixing arises entirely from the charged lepton mass matrix.
Our results apply to any LR theory where neutrino physics below $v_R$ is described by \eqref{eq:LREFT}, together with the gauge interactions. Our results may not apply if there are additional states below $v_R$ (e.g., neutral fermions with bilinear operators mixing with $\nu$ or $N$).

We now consider how the effective theory of (\ref{eq:LREFT}) arises in two simple models. We begin with the conventional LR theory with scalar multiplets $\Delta_L, \Delta_R$ and $\Phi$ which transform as (3,1), (1,3) and (2,2) under $SU(2)_L \times SU(2)_R$, respectively. This leads to the Lagrangian,
\begin{align}
- {\cal L} _{ \rm LR} \, \supset \, y_{ij} \, (\ell_i \,  \bar{\ell}_j) \, \Phi + y'_{ij} ( \ell_i  \, \ell_j ) \, \Delta_L  + y_{ij}'^{(*)} (\bar{\ell}_i  \, \bar{\ell}_j) \, \Delta_R     
  + {\rm h.c.} \,.
\label{eq:convLR}
\end{align}
With this scalar spectrum, the LR symmetry is broken by $\vev{\Delta_R} = v_R$, giving the $ ( N _i N _j ) $ term of \eqref{eq:LREFT}, and $\Phi$ contains the SM Higgs, $H_L$, giving the $( \ell _i  N _j ) $ term of \eqref{eq:LREFT}.  Finally, $\Delta_L$ acquires a mass of order $v_R$ and, when it is integrated out of the theory, leads to the $ ( \ell _i  \ell _j ) $ term of \eqref{eq:LREFT} via the quartic interaction $\lambda_{LR} \,  \Delta_L \Delta_R \Phi^\dagger \Phi$. The constant $c$ is proportional to $\lambda_{LR}$, and hence $c$ is typically of order unity or smaller; $c \gg 1$ requires fine-tuning the mass of $\Delta_L$ to be far below $v_R$ and we do not consider this possibility.

There is a structurally simpler LR model involving just two scalar multiplets $H_L$ and $H_R$ transforming as (2,1) and (1,2) under ${SU(2)}_L \times {SU(2)}_R$.  This theory has the virtue that, if the LR symmetry is taken to include spacetime parity, it solves the strong CP problem~\cite{Babu:1988mw,Babu:1989rb,Hall:2018let}. Furthermore, the vanishing of the SM Higgs quartic coupling at high energies can be understood in this theory from the Higgs Parity mechanism~\cite{Hall:2018let}. The pure doublet symmetry breaking leads to leptonic interactions relevant for neutrino masses above $v_R$ of the form
\begin{align}
- {\cal L} _{ \rm LR} \; \supset   \; f_{ij} \frac{1}{\Lambda} \,(\ell_i \,  \bar{\ell}_j) \, H_L H_R  + f'_{ij} \frac{1}{\Lambda} ( \ell_i  \, \ell_j )\, H_L^2  +  f_{ij}'^{(*)} \frac{1}{\Lambda} (\bar{\ell}_i  \, \bar{\ell}_j) \,H_R^2   
+  {\rm h.c.} \,,
\label{eq:doubLR}
\end{align}
where $\Lambda$ is the UV cutoff for this theory. Inserting the LR symmetry breaking scale, $\vev{H_R} = v_R$, immediately gives \eqref{eq:LREFT}, with $y^{(\prime )}_{ij} = f^{(\prime )}_{ij} v_R/\Lambda$, and the added prediction that $c=1$.

Right-handed neutrino DM in the keV to MeV mass range requires extremely small numbers, whether in the context of (SM + $N$) or a LR theory.  The requirement that $N_1$ is sufficiently light requires (in a LR theory $ y '' = y ' $),
\begin{align}
y''_{11} \; \sim \; \left\{ \begin{array}{lcc} 10^{-20}  \left( \dfrac{M_1}{10 \, \KEV} \right) \left( \dfrac{10 ^{ 15}\, {\rm GeV} }{M _R } \right) & \hphantom{XXXXX} & ( {\rm SM}+N _i ) \\ 10^{-15}  \left( \dfrac{M_1}{10 \, \KEV} \right) \left( \dfrac{10^{10} \, \GEV}{v_R} \right)  &  & ( {\rm LR} ) \end{array} \right.
\label{eq:yi1_0}\,,
\end{align}
where we have normalized $v_R$ to a scale intermediate between the weak and grand unification scales, which will follow from an $N_1$ DM production mechanism studied below.  Right-handed neutrino DM runs counter to the simple seesaw understanding of why the neutrinos are much lighter than the charged fermion masses~\cite{Yanagida:1979as,GellMann:1980vs,Minkowski:1977sc,Mohapatra:1979ia}. In \eqref{eq:SMEFT}, taking $\Lambda \gg 10^{15}$ GeV so that the second term is irrelevant, and taking $M_R \sim 10^{15} \, \GEV$, gives the observed neutrino masses for $y_{ij}$ and $y''_{ij}$ of order unity. Nevertheless, given the exceptionally small numbers that arise in these theories to understand the weak scale ($10^{-32}$) and the cosmological constant ($10^{-120}$), it seems worth pursuing right-handed neutrino DM, especially in LR theories where their existence is a necessity. 

\section{$ N _1 $ stability}

We define $N_1$ as a cosmologically stable right-handed neutrino responsible for the DM density of the universe.
Even though there is no symmetry that stabilizes $N_1$, it may be sufficiently long-lived to be a DM candidate. The dominant decay of $ N _1 $ is driven by $ N _1 - \nu $ mixing controlled by $ y _{ 1i} $; hence $y_{1i} \ll 1$ is needed for $N_1$ to be long-lived.%
\footnote{Note that our numbering of SM neutrinos does not necessarily coincide with the neutrino numbering commonly found in the literature. }

The $N_1-\nu$ mixing angle is given by
\begin{align}
	\sin 2 \theta_1 &\equiv \frac{ v}{M_1} \sqrt{\Sigma_i \; |y_{1i}|^2},
	\label{eq:sin2thetaDef}
\end{align}
where $v\simeq 174$ GeV.
The experimental constraints on $ \sin 2 \theta_1 $ arise from two different processes: For $M_1$ below about $3 ~\KEV$, the dominant constraint on the sterile-active mixing angle comes from overproducing $N_1$ DM via the Dodelson-Widrow mechanism~\cite{Dodelson:1993je}.
For heavier $N_1$, the dominant constraint comes from overproducing photons by $N_1$ DM decays, most prominently through $ N _1 \rightarrow \nu \gamma$~\cite{Adhikari:2016bei}:
\begin{align}
\Gamma_{N_1\rightarrow \nu\gamma} & \simeq \frac{9 \alpha}{8192 \pi^4}  \; \frac{M_1^5}{v^4} \; \sin^2 2 \theta_1 \,,\notag \\ 
& \simeq  \left(1.5 \times 10^{30} \sec \right)^{-1} \left(\frac{M_1}{1 ~\KEV}\right)^5 \left( \frac{ \sin^2 2 \theta_1 }{5 \times 10 ^{ - 9} } \right) \,.
\label{eq:mixing}
\end{align}
As the decay rate is $ \propto M _1 ^5 $ it grows rapidly with $ M _1 $ and is a powerful constraint on the mixing angle for $ M _1 \gtrsim  {\rm keV}  $. Sufficient stability of $N_1$ requires $\Gamma_{N_1 \rightarrow \nu \gamma} \lesssim 1 \times 10^{-27} s^{-1}$ \cite{Adhikari:2016bei} and hence 
\begin{align}
|y_{1i}| \; \lesssim  \; 10^{-13}  \left( \frac{10 \, \KEV}{M_1} \right)^{3/2} \qquad (M_1 \gtrsim 3~ {\rm keV} ) \,.
\label{eq:yi1}
\end{align}
The combination of the constraints leads to a limit on the mixing angle~\cite{Adhikari:2016bei},
\begin{equation}
\sin ^2 2 \theta_1 \leq 5 \times 10^{-9} 
\begin{cases}
\left(\dfrac{M_1}{3 ~\KEV}\right)^{-1.8} \times D & 
\quad \text{(Overproduction)}\\
\left(\dfrac{M_1}{3 ~\KEV}\right)^{-5} & 
\quad \text{(Decay)}.
\end{cases}
\label{eq:sin2theta}
\end{equation}
$D$ is the dilution factor required to reduce a thermal yield of $N_1$ to the correct DM abundance.  The higher photometric sensitivities of next generation x-ray and gamma-ray telescopes such as ATHENA~\cite{Nandra:2013shg} and  e-ASTROGAM~\cite{Tatischeff:2016ykb} may probe an order of magnitude smaller $\sin^2 2\theta_1$~\cite{Caputo:2019djj}. For $M_1 > 1$ MeV, the tree-level decay $N_1 \rightarrow e^+ e^- \nu$ is open and the resultant constraint on $y_{1i}$ is similar to (\ref{eq:yi1}).

The smallness of the Yukawa coupling in \eqref{eq:yi1} can be explained in the SM+$ N _i $ theory by imposing a discrete $Z_2$ symmetry under which $N_i$ are odd so that the last operator of \eqref{eq:SMEFT} is forbidden, giving $y_{1i} =0$ and making $N_1$ stable.  Furthermore, introducing an inert doublet $H'_L$, that has no vacuum expectation value and is odd under this parity, allows the first operator of (\ref{eq:SMEFT}) to be generated from a 1-loop radiative correction~\cite{Ma:2006km}.  In the LR framework, a $Z_2$ symmetry that sets $y_{1i}=0$ also forbids charged lepton Yukawa couplings.  However, in a LR theory one can alternatively impose a discrete $Z_{4L} \times Z_{4R}$ symmetry setting $y_{1i}=0$, guaranteeing cosmologically stable $N_1$, while allowing charged lepton masses. We discuss how the model works in Appendix \ref{sec:N1stab}.

If kinematically allowed, $N_1$ can also beta decay via $W_R$ exchange to $\ell^\pm +$ hadron(s), where $\ell^\pm$ is any charged lepton,
regardless of how small $y_{1i}$ is. The inclusive decay rate is,
\begin{align}
\Gamma_{N_1 \rightarrow \ell^\pm +{\rm hadrons} } &\simeq \frac{3}{1536 \pi^3} \frac{M_1^5}{v_R^4} \nonumber \\
&\simeq \left(1.4 \times 10^{24} \sec\right)^{-1} \left(\frac{M_1}{150 ~\MEV}\right)^5 \left(\frac{v_R}{10^{10} ~\GEV}\right)^{-4} \,.
\label{eq:WRdecay}
\end{align} 
For sufficiently small $M_1$ or large $v_R$, this is below the observational upper bounds of $\sim 10^{25} \sec $~\cite{Essig:2013goa}.%
\footnote{Although $N_1$ decays via $W_R$ to charged pions which then decay to muons, for $M_1 \gg m_\pi$, there may be a large number of neutral pions in the decay shower, which subsequently decay to hard photons and yield slightly stronger constraints on the $N_1$ lifetime~\cite{Essig:2013goa}.}
Here the decay rate is estimated in the quark picture, but the interpolation to the meson regime $M_1 \gtrsim  m_\pi$ is correct at the order of magnitude level. 

For $M_1$ below the pion mass, beta decay to $\ell^+ \ell^- \nu$ via $W_R - W_L$ mixing is important. This decay channel is also independent of $y_{1i}$ and given by
\begin{align}
    \Gamma_{N_1 \rightarrow \ell^+ \ell^- \nu} \simeq 
    \frac{\Gamma_{N_1 \rightarrow \ell^\pm +{\rm hadrons}}}{3} \times 
    \begin{dcases}
     1 &\quad :\text{(2,2) Breaking} \\
        \left(\frac{1}{16\pi^2}\frac{m_b m_t}{v^2}  \ln\left(\frac{\Lambda}{v} \right)\right)^2 &\quad : \text{(2,1)+(1,2) Breaking}.
    \end{dcases}
    \label{eq:N1WRWL1}
\end{align}
When the electroweak symmetry is broken by an $SU(2)_L\times SU(2)_R$ bi-fundamental scalar, its vacuum expectation value gives a $W_R-W_L$ mixing at tree-level. If the electroweak symmetry is broken by an $SU(2)_L$ doublet scalar, the mixing is generated by a top/bottom quark loop. The quantum correction is logarithmically divergent in the effective theory where the quark masses are given by dimension-5 operators, similar to Eq.~ (\ref{eq:doubLR}). The scale $\Lambda$ that cuts off the divergence is model-dependent, but is close to $v_R$ since the top and bottom Yukawa couplings are not small.

The $W_R-W_L$ mixing also induces the decay of $N_1$ into $\nu\gamma$~\cite{Lavoura:2003xp,Bezrukov:2009th,Greljo:2018ogz}. Connecting the $\ell^+ \ell^-$ in the beta-decay diagram and attaching an external photon to this loop gives a decay rate
\begin{align}
    \Gamma_{N_1\rightarrow \nu\gamma} \simeq  \frac{\alpha}{4\pi}\frac{m_{\tau}^2}{M_1^2} \Gamma_{N_1 \rightarrow \ell^+ \ell^- \nu},
    \label{eq:N1WRWL2}
\end{align}
which is observationally limited by photon searches to be less than about $10^{-27} s^{-1}$.

\section{Relativistic freeze-out and dilution}
\label{sec:cosmology}
The right-handed neutrinos couple to the SM bath via $W_R$ exchange. If the reheat temperature of the universe after inflation is sufficiently high,
\begin{align}
\label{eq:eqtemp}
T_{\rm{RH}} ^{ {\rm inf}} \gtrsim 10^8 ~\GEV \left(\frac{v_R}{10^{10} ~\GEV}\right)^{4/3},
\end{align}
the right-handed neutrinos reach thermal equilibrium and subsequently decouple with a thermal yield $Y_{\rm{th}} \simeq 0.004$.%
\footnote{The analysis is this section is also applicable to lower $T_{\rm RH}^{\rm inf}$ 
as long as $N_1$ and $N_2$ are frozen-in from $W_R$ exchange, and $N_1$ is overproduced as DM (see Eq.~\eqref{eq:FIAbundance}). In such a scenario, the required dilution to realize $N_1$ DM is diminished, and hence the warmness constraints on $N_1$ slightly increase above $2 \, \KEV$. See Fig.~\ref{fig:thermal_FI} for the warmness constraints on a pure freeze-in cosmology without any dilution.}
For $N_1$ to have the observed DM abundance requires $m_{N_1} \simeq 100$ eV; however, such light sterile neutrino DM is excluded by the Tremaine-Gunn~\cite{Tremaine:1979we,Boyarsky:2008ju,Gorbunov:2008ka} and warmness~\cite{Narayanan:2000tp,Irsic:2017ixq,Yeche:2017upn,Seljak:2006qw} bounds; see~\cite{Adhikari:2016bei} for a recent review.

Nevertheless, it is still possible to realize $N_1$ as DM with $M_1 \gtrsim \KEV $, if they decouple relativistically from the thermal bath and their abundance is diluted. If another right-handed neutrino, $N_2$, is sufficiently long-lived such that it comes to dominate the energy density of the universe and produces entropy when it decays, it can dilute the DM abundance and cool $N_1$ below warmness bounds~\cite{Asaka:2006ek,Bezrukov:2009th}. 
The relic density of $N_1$ is
\begin{align}
\frac{\rho_{N_1}}{s} & =  1.6 \; \frac{3}{4} \; \frac{ M_1}{M_2} \; T_{\rm RH} \,, \nonumber \\ 
\Rightarrow \frac{ \Omega_{N_1} }{ \Omega _{ {\rm DM}}} & \simeq \left( \frac{ M _1 }{ 10 \, {\rm keV} } \right) \left( \frac{ 300 \,{\rm GeV} }{ M _2 } \right) \left( \frac{ T _{\rm RH}}{ 10 \, {\rm MeV}  } \right) \,,
\label{eq:DMabund}
\end{align}
where the numerical factor $1.6$ is taken from~\cite{Harigaya:2018ooc}, $\rho _{ N _1 } $ is the energy density, $ s $ is the entropy density, $ \Omega _{ {\rm DM}} \simeq 0.25 $ is the observed cosmic relic abundance, and $T_{\rm RH}$ is the decay temperature of $N_2$, as set by its total decay rate $\Gamma_{N _2 }$
\begin{align}
T_{\rm RH} = \left(\frac{10}{\pi^2 g_*}\right)^{1/4}\sqrt{\Gamma_{ N _2 } \mpl}.
\label{eq:TRH}
\end{align}
These formulae are also applicable to the case where $N_3$ first dominates the universe and decays to create entropy, and later $N_2$ dominates and creates entropy again.
Inserting the warmness bound on $N_1$, $(M_1 > 2 \, \KEV$, see Sec.~\ref{sec:warmness}), and the reheating bound from hadronic decays of $N_2$ during BBN ($T_{\rm RH} > 4 \,\MEV$)~\cite{Kawasaki:1999na,Kawasaki:2000en,Hasegawa:2019jsa},%
\footnote{Low reheating temperatures can also affect the CMB since some decays occur after neutrinos decouple, reducing the effective number of neutrinos~\cite{Kawasaki:1999na,Kawasaki:2000en,Ichikawa:2005vw}. In our case, $N_2$ also decays into neutrinos and the bound from the CMB, $T_{\rm RH}> 4$ MeV~\cite{deSalas:2015glj}, may be relaxed.}
into~\eqref{eq:DMabund} requires%
\footnote{Ref.~\cite{Nemevsek:2012cd} points out that if the mass eigenstate $N_1$ forms an $SU(2)_R$ doublet with the mass eigenstate $\tau$, it might be possible for $N_1$ to decouple earlier than $N_2$ because of the Boltzmann-suppressed density of $\tau$ relative to $\mu$ or $e$. This reduces the relic density of $N_1$ compared to $N_2$ which relaxes the necessary dilution from $N_2$ by a factor of $3-4$ and hence lowers the bound on $M_2$ from $24 \, \GEV$ to $6-8 \, \GEV$. However, if $N_2$ decouples after $N_1$, its density is Boltzmann-suppressed since $M_2 \gg m_\tau$, making dilution ineffective. Consequently, we find that the potential relaxation of the bound on $M_2$ \eqref{eq:m2bound} unattainable. Ref.~\cite{Nemevsek:2012cd} also points out that if $N_2$ forms an $SU(2)_R$ doublet with $\mu$ and $M_2 \simeq m_\mu + m_\pi \simeq \, 250 \, \MEV$, the decay rate of $N_2$ via $W_R$ exchange is suppressed by a small phase space, allowing $N_2$ to be long-lived and provide sufficient dilution even for $v_R$ around the TeV scale. Since we find that the possible relaxation of the lower bound on $M_2$ does not work, we also cannot confirm this claim.}
\begin{align}
	M_2 \gtrsim 24 \, \GeV.  \hspace{1cm} &  (\text{Warmness, reheating, DM abundance})\,.
	\label{eq:m2bound}
\end{align}

There are several possible decay modes for $N_2$, and which one dominates varies with $M_2$. $N_2$ can always beta decay through $W_R$ exchange into right-handed fermions, $N_2 \rightarrow (\ell^+ \bar{u} d,\, \ell^- u \bar{d})  $ and $N_2 \rightarrow N_1 \ell^+ \ell^- $. These decay channels are unavoidable as they are independent of the free-parameter $y_{2i}$, and prevent $N_2$ from efficiently diluting $N_1$ in some regions of parameter space. The $N_2$ decay rate via $W_R$ exchange is 
\begin{align}
\Gamma_{N _2 \rightarrow N _1 \ell^+ \ell^-} + \Gamma_{N _2 \rightarrow  (\ell^+ \bar{u} d, \, \ell^- u \bar{d})}= \frac{1}{1536\pi^3} \frac{M_2^5}{v_R^4} \times 20\,.
\label{eq:2beta}
\end{align}
In addition, when $M_2 \gtrsim v$, $N_2$ can decay at tree-level via $ N _2 \rightarrow \nu  h ,  \nu Z , \ell ^\pm  W ^\mp $ while for $M_2 \lesssim v$, $N_2$ can beta decay through $W_L/Z$ exchange and active-sterile mixing to SM fermions, $N_2 \rightarrow \ell ud, \ell^+ \ell^- \nu, \nu \nu \bar{\nu}$. 
These decay rates are given by
\begin{align}
\Gamma_{N _2 \rightarrow \ell H_L} & = \frac{1}{8 \pi} \sum_i |y_{2i}|^2 M_2  \hspace{-2cm} & (M_2 \gtrsim v)
\label{eq:ellH} \\
\Gamma_{N_2 \rightarrow (\ell^+ \bar{u} d, \,\ell^+ \ell^-\bar{\nu}, \, \nu \nu \bar{\nu}\,\, \text{or h.c.})}
& \simeq \frac{171}{8}\frac{1}{1536\pi^3} \frac{M_2^3}{v^2} \sum_i |y_{2i}|^2  \hspace{-2cm}&  (M_2 \lesssim v).
\label{eq:ellud}
\end{align}
For the latter we add up the results in~\cite{Gorbunov:2007ak,Bezrukov:2009th} in the limit of a vanishing Weinberg angle, for simplicity.
In either case, $y_{2i}$ must be sufficiently small so that $N_2$ dominates the energy density of the universe before decaying. Diluting $N_1$ to the observed DM abundance requires 
\begin{equation}
|y_{2i}| \lesssim  
\begin{cases}
3\times 10^{-10}\left(\dfrac{M_2}{24 ~\GEV}\right)^{-1/2}\left(\dfrac{M_1}{2 ~\KEV}\right)^{-1} & (M_2 \lesssim v)\\
1\times 10^{-11}\left(\dfrac{M_2}{v}\right)^{1/2}\left(\dfrac{M_1}{2 ~\KEV}\right)^{-1} & (M_2 \gtrsim v).
\end{cases}
\label{eq:yi2}
\end{equation}
The equality sign applies when the contribution to the $N_2$ decay rate from $W_R$ exchange, (\ref{eq:2beta}), is sub-dominant. 

In Appendix \ref{sec:nuMasses}, we use the above results, together with the radiative stability bound on $N_1$, to derive constrains on the neutrino mass matrix of \eqref{eq:numassmatrix}. 

\begin{enumerate}
    \item For $M_1 < M_3$, we show that the lightest neutrino mass eigenstate is closely aligned with $\nu_1$ and has a mass $m_1 \ll \sqrt{\smash[b]{ \Delta m_{\rm sol}  ^2}}$.  In the case that $M_3 > M_2$, the other two mass eigenstates are very close to $\nu_2$ and $\nu_3$ and have masses $m_2 = c \, (v^2/v_R^2) M_2$ and $m_3 = c \,(v^2/v_R^2) M_3 - y_{33}^2 v^2/M_3$. For $M_3 < M_2$, the $(2,3)$ entry of the mass matrix may be non-negligible, so that the two heavy active mass eigenstates are each linear combinations of $\nu_{2,3}$.  In this case, we are able to derive a relation between the scale of their masses and $M_2$: $M_2 \simeq \mu (v_R/v)^2 c^{-1}$, \rm{where} $0.01 \, \EV \lesssim \mu \lesssim 0.10 \, \EV$. In the rest of the paper we take
\begin{align}
M_2 &\simeq m_2 \left(\frac{v_R}{v}\right)^2 \frac{1}{c},
\label{eq:M2m2}
\end{align}
but, in the case that $M_3 < M_2$, $m_2$ should be taken in the range $(0.01 \, - 0.10) \, \EV$ and not set to an active neutrino mass eigenvalue.
\item For $M_1>M_3$, we show that the lightest neutrino mass is much smaller than $\sqrt{\smash[b]{ \Delta m_{\rm sol}  ^2}}$ and that $M_2$ is given by Eq.~(\ref{eq:M2m2}), with the parameter $c$ replaced by a parameter $c_{\rm eff} < c$.
 \end{enumerate}

\begin{figure}[tb]
    \centering
    \begin{minipage}{0.5\textwidth}
        \centering
        \includegraphics[width=1\textwidth]{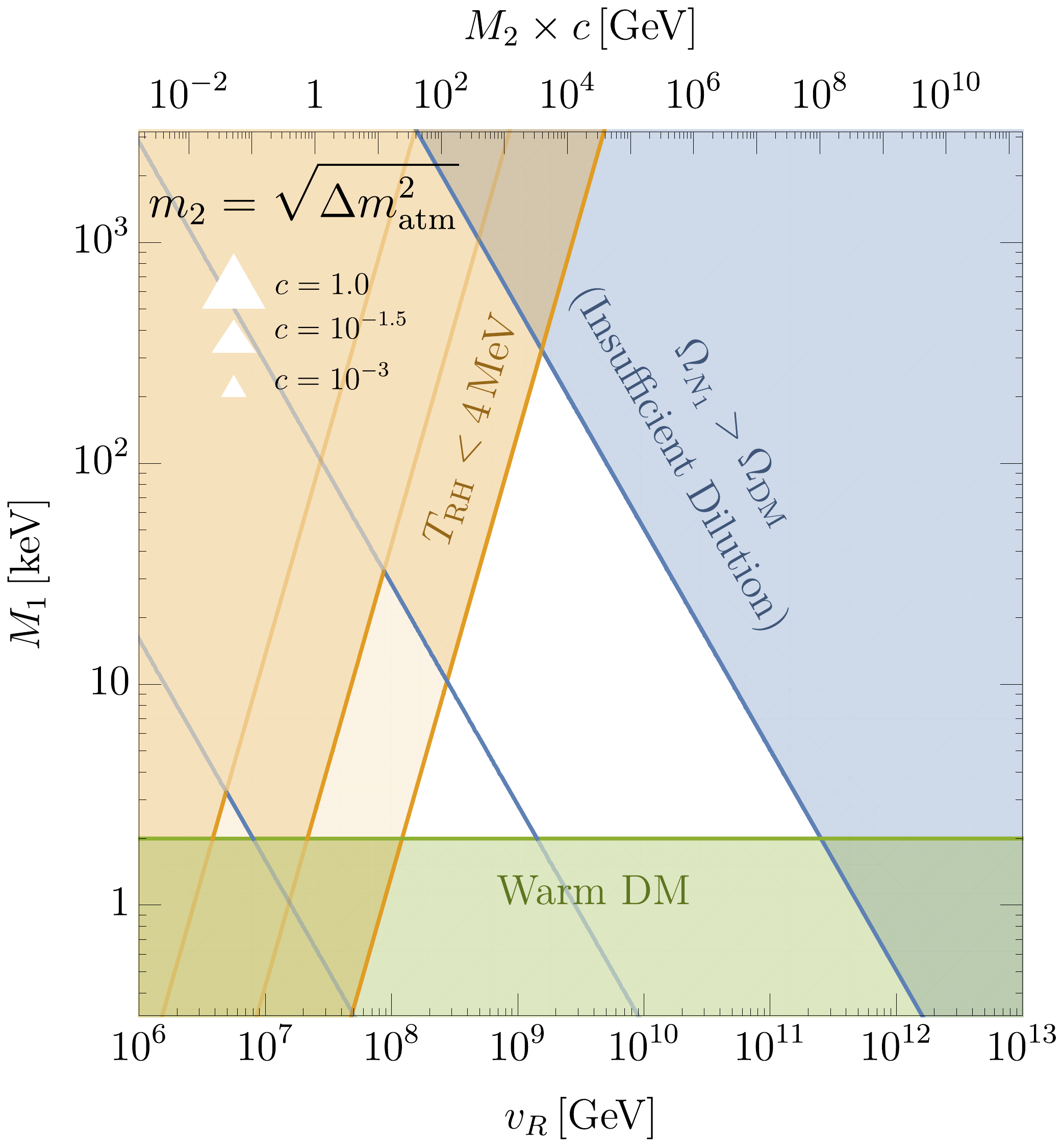} 
    \end{minipage}\hfill
    \begin{minipage}{0.5\textwidth}
        \centering
        \includegraphics[width=1\textwidth]{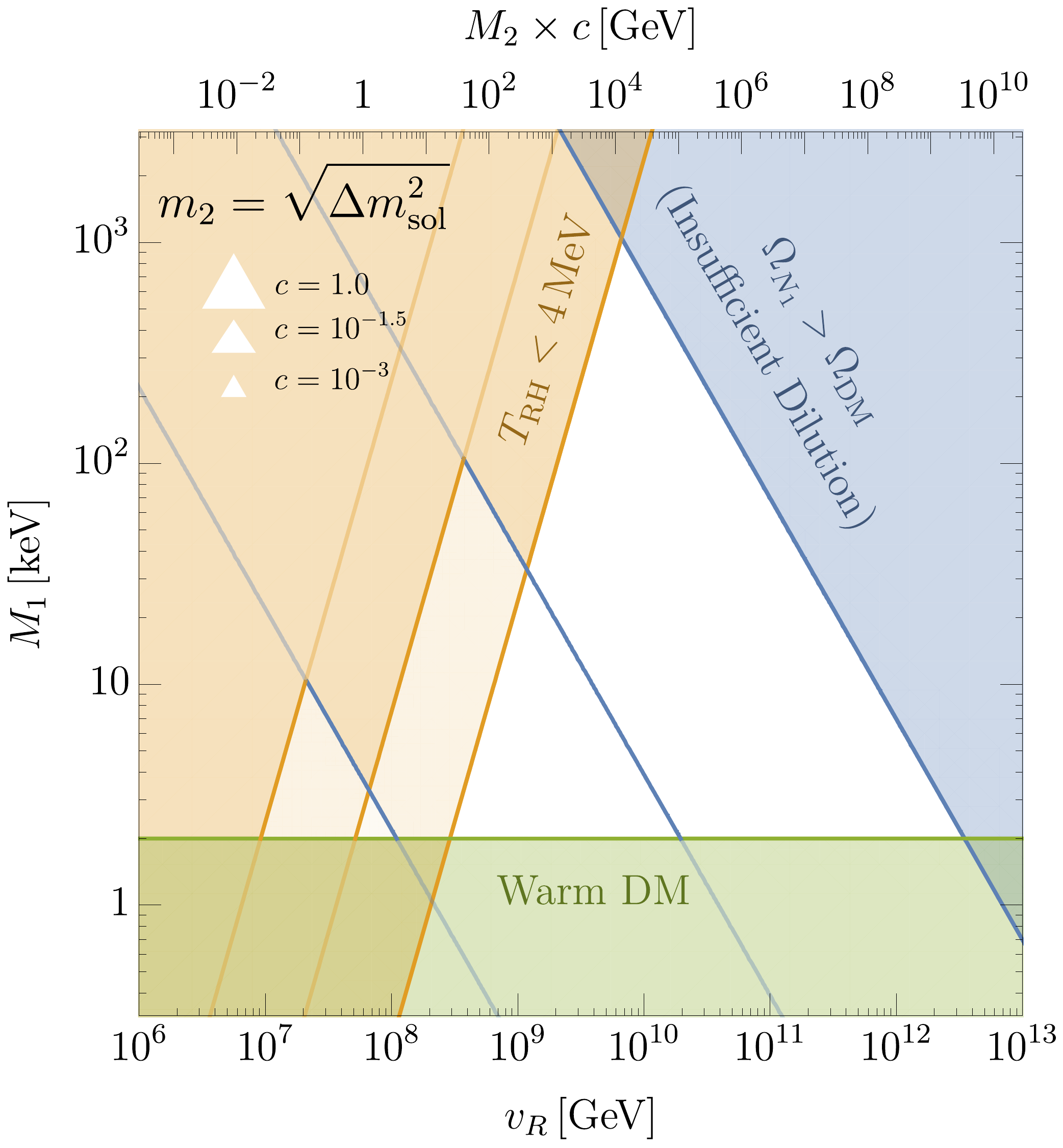} 
    \end{minipage} 
    \caption{The parameter space of $N_1$ DM produced by relativistic freeze-out and dilution from $N_2$ decay: constraints on the LR symmetry breaking scale $v_R$ and the mass $N_1$. The constraints from warm DM are in {\color{c3}\bf green}, Big Bang Nucleosynthesis in {\bf \color{c2}  orange}, and insufficient dilution in {\bf \color{c1} blue}. The constraints depend on the LR-model dependent parameter $c\lesssim 1$. {\bf Left}: We fix the $\nu_2$ mass by the atmospheric neutrino mass difference, $m_2 = \sqrt{ \smash[b]{ \Delta m_{\rm atm}  ^2}}$. {\bf Right}: We fix the $\nu_2$ mass by the solar neutrino mass difference, $m_2 = \sqrt{ \smash[b]{ \Delta m_{\rm sol}  ^2}}$.}
    \label{fig:thermal_FO}
\end{figure}

In Fig.~\ref{fig:thermal_FO}, we show the constraints on $(v_R,M_1)$ when $m_2 = \sqrt{ \smash[b]{ \Delta m_{\rm atm}  ^2}} $ $\,$({\bf left}) and $m_2 = \sqrt{\smash[b]{ \Delta m_{\rm sol}  ^2}}$ $\,$({\bf right}). 
In the orange shaded region, the required $T_{\rm RH}$ is below 4 MeV, which is excluded by hadronic decays of $N_2$ during BBN~\cite{Kawasaki:1999na,Kawasaki:2000en}. 
The green-shaded region is excluded due to the warmness of $N_1$ affecting large scale structure. To the right of the blue line, the beta decay rate of $N_2$ via $W_R$ exchange is the dominant contribution to $\Gamma_{ N _2 }$; here, the dilution of $N_1$ is chiefly through $N _2 \rightarrow N _1 \ell^+ \ell^-$ and $N_2 \rightarrow (\ell^+ \bar{u} d, \, \ell^- u \bar{d})$. Using (\ref{eq:M2m2}), these decay rates scale as a positive power of $ M _2 $ and hence $ v _R $. Within the blue-shaded region, the $ N _2 $ decay rate becomes too fast to efficiently dilute the $ N _1 $ energy density.

The blue line itself is an interesting region of parameter space, which does not require any tuning but simply corresponds to the limit where the the dominant decay is set entirely by the $ W _R $ exchange terms in (\ref{eq:2beta}). In this limit the $ N _1 $ abundance has two contributions: from $ N _2 $ decay through $ N _2 \rightarrow  N _1 \ell^+ \ell^- $ as well as the abundance from relativistic decoupling. While the latter is the dominant component, the former can also make up a significant component of DM, which can be probed by future experiments as discussed in Sec.~\ref{sec:hotness}.

As can be seen from Fig.~\ref{fig:thermal_FO}, the allowed region of frozen-out $N_1$ DM from LR theories forms a bounded triangle in the $v_R-M_1$ plane. The position and size of the triangle depends on $c$, such that the allowed region shrinks in size and shifts to lower $v_R$ for smaller $c$. This is because the $\Omega_{N_1} > \Omega_{\rm DM}$ bound depends more sensitively on $M_2$ (and hence $c$) than the $T_{\rm RH} < 4 ~\MEV$ bound.  We show the effect of $c$ on the allowed region for three values of $c$: one near the experimental minimum, one near the natural maximum, and one in between. As can be seen by the smallest triangle of Fig.~\ref{fig:thermal_FO}, the allowed region of $N_1$ DM disappears for $c \lesssim 1 \times 10^{-4}$, placing an experimental lower bound on $v_R \gtrsim 10^6 ~\GEV$. Similarly, the naturalness argument discussed in Sec.~\ref{sec:nuinLR} limits $c$ near unity, and an upper bound on $v_R \lesssim 10^{13} ~\GEV$ as shown by the largest triangle of Fig.~\ref{fig:thermal_FO}. For the remainder of this paper, we conservatively focus on the case $c = 1$, the largest naturally allowed parameter space of $N_1$ DM, when considering signals and future experimental probes.

\section{Signals and future probes}
So far we have focused on the current constraints on sterile neutrino DM in general LR theories and found freeze-out to be a viable option as long as the $c$ parameter, characterizing the seesaw contribution to the light neutrino masses, is not too small.
In this section, we discuss how future observations can probe the parameter space through dark radiation, warm DM, and additional structure on very small scales.
In addition, the requirement of viability of leptogenesis greatly restricts the parameter space.

\label{sec:signals}
\begin{figure}[]
    \centering
    \begin{minipage}{0.5\textwidth}
        \centering
        \includegraphics[width=1\textwidth]{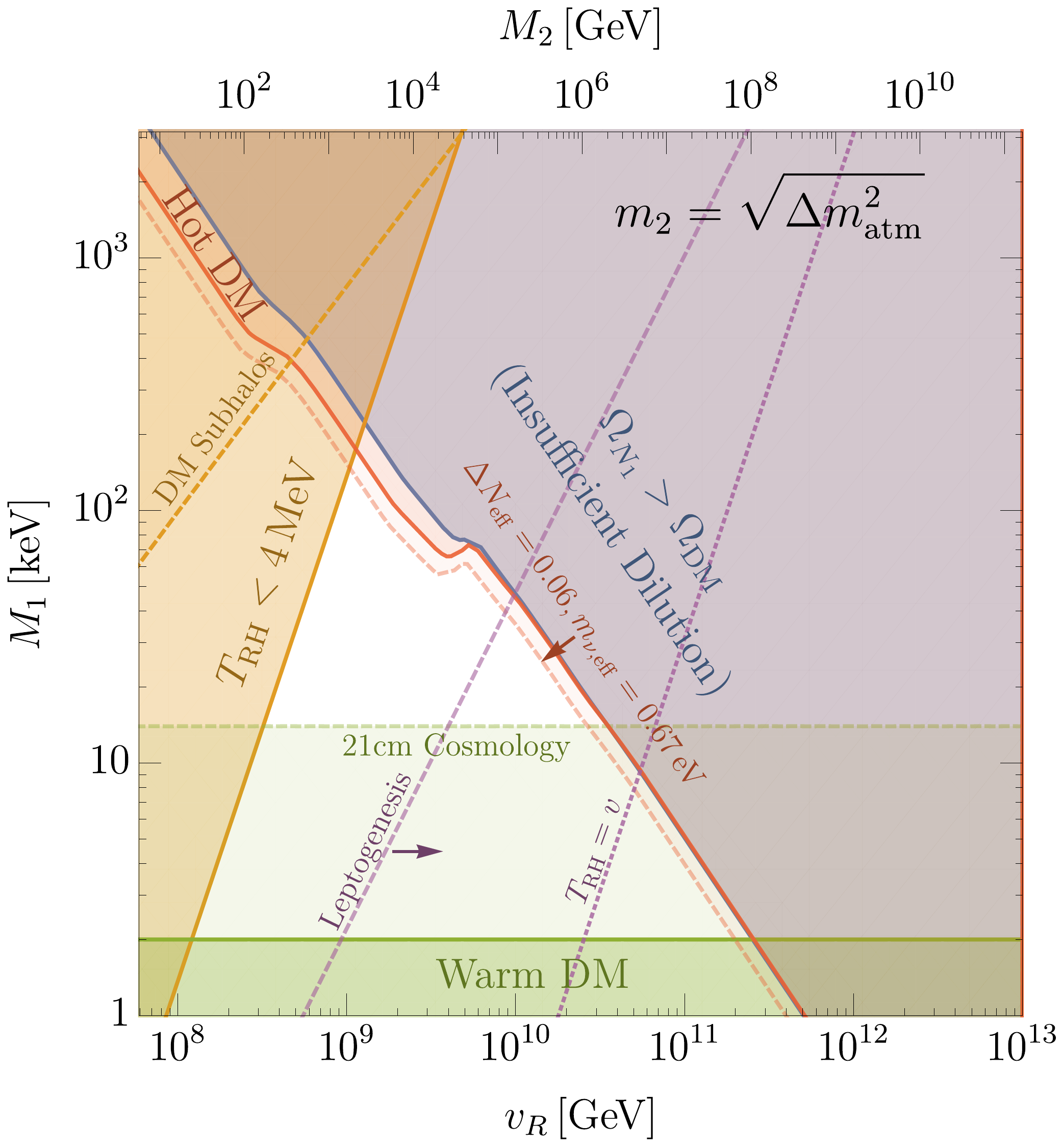} 
    \end{minipage}\hfill
    \begin{minipage}{0.5\textwidth}
        \centering
        \includegraphics[width=1\textwidth]{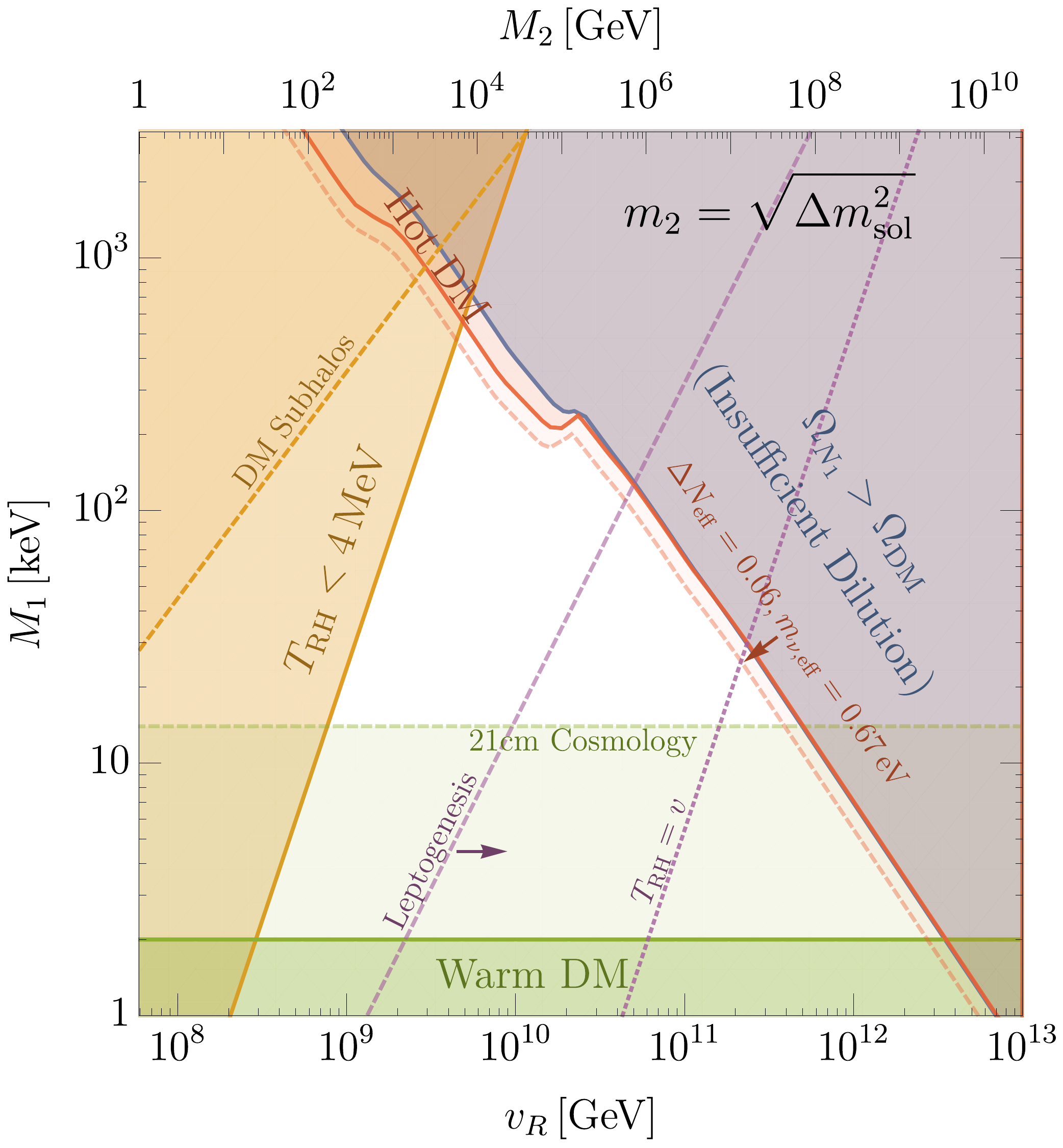} 
    \end{minipage}
    \caption{The parameter space of $N_1$ DM produced by relativistic freeze-out and dilution from $N_2$ decay in terms of the left-right symmetry breaking scale, $ v _R $, and the mass of $ N _1 $, $ M _{ 1 } $, for $c = 1$. We show constraints from $ N _2 $ decaying after Big Bang Nucleosynthesis ({\color{c2} \bf orange}), decaying too early to provide sufficient $ N _1 $ dilution ({\color{c1} \bf blue}), warm DM bounds ({\color{c3} \bf green}), and hot DM bounds ({\color{c4} \bf red}). In addition we show prospects of future surveys of $T_{\rm RH}$ from pulsar timing on DM subhalos (dashed {\color{c2} \bf orange}), improved searches for hot DM from CMB telescopes (dashed {\color{c4} \bf red}), and warm DM from $21$-cm cosmology (dashed {\color{c3} \bf green}). Lastly, to the left of the dashed {\color{c5} \bf purple} curve labeled `Leptogenesis', the baryon asymmetry produced by $N_2$ decays is insufficient due to dilution and sphalerons, even with $\epsilon = 1$.
    {\bf Left:} We fix the $ \nu _2 $ mass with the atmospheric neutrino mass difference, $m_2 = \sqrt{ \smash[b]{ \Delta m_{\rm atm}  ^2}}$. {\bf  Right:} We fix the $ \nu _2 $ mass with the solar neutrino mass difference, $m_2 = \sqrt{ \smash[b]{ \Delta m_{\rm sol}  ^2}}$.}
    \label{fig:signals}
\end{figure}

\subsection{Warmness}
\label{sec:warmness}
The free-streaming length of thermally produced $N_1$ can be large if $N_1$ is light. When $M_1$ is $\mathcal{O(\KEV)}$, the free-streaming length of $N_1$ approaches the size of galactic mass perturbations, suppressing the matter power spectrum on scales $k \gtrsim 0.1~{\rm Mpc}^{-1}$~\cite{Bond:1980ha,10.1143/PTP.64.2029,Klinkhamer:1981uy,Wasserman:1981uz,Olive:1981ak}. A suppression can be observed through large scale structure surveys, perturbations in the cosmic microwave background (CMB), or absorption of low-redshift Lyman-$\alpha$ photons by neutral hydrogen (a tracer of DM) in the intergalactic medium~\cite{McDonald:2004eu,Viel:2013apy,Aghanim:2018eyx,Banik:2018pjp}. For $N_1$ that was thermally produced and diluted to the observed DM abundance, the bounds are at $ {\cal O} ( 1-5~ {\rm keV} ) $ range. We adopt $M_1 \gtrsim  2 ~ {\rm keV} $ as our constraint in the dark green region of Fig.~\ref{fig:signals}. Future $21$-cm cosmology experiments, which can trace early star and galaxy formation at cosmic dawn, are anticipated to probe the matter power spectrum on scales $k \gtrsim 50~{\rm Mpc}^{-1}$. If no suppression on such scales is observed, searches would constrain $M_1 \gtrsim 14~ {\rm keV}  $~\cite{Munoz:2019hjh}, which we show with the dashed green region of Fig.~\ref{fig:signals}.

\subsection{Hotness}
\label{sec:hotness}
Although $N_1$ DM is dominantly produced thermally, a subdominant fraction is always produced non-thermally (see section~\ref{sec:cosmology}). Specifically, the beta decay $N_2 \rightarrow N_1 \ell^+ \ell^-$ produces relativistic $N_1$. This non-thermal population of $N_1$ becomes non-relativistic at temperatures $ {\cal O} (  {\rm eV})  $ and contributes a hot component of DM. Constraints on hot DM are conventionally given in terms of the effective number of neutrino species, $\Delta N_{\rm eff}$, which parameterizes its energy density while relativistic, and the effective neutrino mass, $m_{\nu,{\rm eff}}$, which parameterizes its energy density when it has become non-relativistic matter~\cite{Aghanim:2018eyx,Feng:2017nss}. For LR models, this is given by
\begin{align}
\Delta N_{\rm eff} & \; = \; \frac{1}{3} {\rm Br}(N_2 \rightarrow N_1 \ell^+ \ell^-) \left(\frac{g_{*,\rm{eq}}^4}{g_{*,\rm{T_{\rm RH}}}}\right)^{\tfrac{1}{3}} \frac{4}{7}\left(\frac{4}{11}\right)^{-\tfrac{4}{3}} \simeq \;  0.97 \; {\rm Br}(N_2 \rightarrow N_1 \ell^+ \ell^-)\left(\frac{106.75}{g_{*,\rm T_{\rm RH}}}\right)^{\tfrac{1}{3}}, \nonumber\\
m_{\nu,{\rm eff}} & \; \equiv  \; 94.1 \, {\rm eV} \; \Omega_{N_1,{\rm hot}} h^2 \; \simeq  \; 11~{\rm eV} \; {\rm Br}(N_2 \rightarrow N_1 \ell^+ \ell^-).
\end{align}
When the $ W _R $-exchange decay is subdominant, Br$(N_2 \rightarrow N_1 \ell^+ \ell^-)$ scales as $ M _1 ^2 v _R ^2 $, and it saturates at $0.1$ (since 90\% of beta decays produce quarks and no $N_1$) along the blue curve.
Along this line a significant amount of hot DM is predicted: $\Delta N_{\rm eff} \simeq 0.1$ and $m_{\nu,{\rm eff}} \simeq 1.1 \,\EV$. Coincidentally, current limits on the two-dimensional marginalized distribution of $\Delta N_{\rm eff}$ and $m_{\nu,{\rm eff}}$ already require $\Delta N_{\rm eff} \lesssim 0.1$ and $m_{\nu,{\rm eff}} \lesssim 1.0 \,\EV$~\cite{Feng:2017nss}, which we indicate by the red-shaded region labeled `Hot DM' in Fig.~\ref{fig:signals}.

CMB Stage IV~\cite{Abazajian:2019eic}, a collection of future ground based telescopes, will be able to search for hot DM signals inside the currently allowed region. Assuming a null detection, the experiment will be able limit $\Delta N_{\rm eff} \lesssim  0.06$~\cite{Abazajian:2016yjj}, which we show by the dashed red region of Fig.~\ref{fig:signals}.%
\footnote{Future space based telescopes such as CORE can theoretically detect $m_{\nu,{\rm eff}} \sim 0.04 \EV$ at $1\sigma$, but only if $\Delta N_{\rm eff} \gtrsim 0.05$~\cite{Boser:2019rta,DiValentino:2016foa}.}
Note that for $v_R \lesssim 10^{10} \, \GEV$, $T_{\rm RH}$ occurs below the QCD phase transition, which is accompanied by a sharp decrease of $g_{*,T_{\rm RH}}$, leading to an enhancement in $ \Delta N _{ {\rm eff}} $ and strengthening the red-shaded region. 
The limit where the the dominant decay of $N_2$ is set by the $W_R$ exchange can be probed by CMB Stage IV.

\subsection{Early matter dominated era}
The current bound on the reheat temperature, $T_{\rm RH} \gtrsim  4 ~ \MEV$, comes from $N_2$ decaying during BBN, leading to its decay products altering the neutron to proton ratio
enough to conflict with the observed light element abundances~\cite{Kawasaki:1999na,Kawasaki:2000en}. Presently, ideas to probe higher reheat temperatures rely on the cosmological effects of the early matter dominated era, namely the formation of ultra compact DM halos~\cite{Erickcek:2011us,Choi:2017ncz}. For example, halos with masses as low as $M_{\rm halo} \simeq 10^{-10} M_{\odot}$ can be observed with pulsar timing arrays once the Square Kilometer Area~\cite{Keane:2014vja} is built~\cite{Dror:2019twh} (in principle, gravitational microlensing could also be used to look for sub-halos from early matter domination, but such halos would typically have concentration parameters of $ {\cal O} ( 10 ^3 ) $ and would be too diffuse to have a sizable signature~\cite{Dror:2019twh,Croon:2020wpr,Bai:2020jfm}). The largest DM halo masses are correlated with $T_{\rm RH}$
since the density perturbations, $k(a) \equiv  a H(a)$, which enter the horizon during the early matter dominated era and source the halos, are largest just before reheating: 
\begin{align}
    M_{\rm halo} \approx  \frac{4}{3}\pi k_{\rm RH}^{-3} \rho_{m,0} \approx  10^{-10} M_{\odot} \left(\frac{T_{2}}{500 \,\MEV}\right)^{-3} \left(\frac{g_{*s}(T_{2})}{68}\right)  \left(\frac{g_{*}(T_{2})}{68}\right)^{-3/2}. \label{eq:mHaloTRH}
\end{align}
Here, $k_{\rm RH} = a(T_{\rm RH})H(T_{\rm RH})$ is the scale of density perturbations entering the horizon at $T_{\rm RH}$, and $\rho_{m,0}$ is the present-day mass density of non-relativistic matter~\cite{Erickcek:2011us}. From \eqref{eq:mHaloTRH}, we see that pulsar timing arrays can probe reheat temperatures as high as $\sim 500 \, \MEV$. 

An important caveat to these experimental searches arises when DM has such a large free-streaming length that ultra compact halos cannot form during the early matter-dominated era. The free-streaming length of $N_1$ DM is~\cite{Kolb:1990vq}
\begin{align}
    \lambda_{\rm FS} &\equiv \int _0^{t_{\rm eq}} \frac{v(a)}{a} dt \leqq \int _{t_{\rm RH}}^{t_{\rm eq}} \frac{v(a)}{a} dt = \frac{1}{H_{\rm RH} a_{\rm RH}^2}\frac{\langle p_{\rm dec}\rangle a_{\rm dec}}{M_1} \ln\left(\frac{h(a_{\rm eq})}{h(a_{\rm RH})} \right),
\end{align}
where $h(a) \equiv \sqrt{a^2 + (\langle p_{\rm dec}\rangle a_{\rm dec}/M_1)^2} + a $ and,
\begin{align} 
\langle p_{\rm dec}\rangle &\simeq 3.2 \, T_{\rm eq}\frac{a_{\rm eq}}{a_{\rm dec}}\left( \frac{g_{*s, {\rm eq}}}{g_{*s,{\rm dec}}}\frac{\rho_{\rm DM}/s}{M_1 Y_{\rm therm}}\right) ^{1/3} 
\end{align} 
is the average momentum of $N_1$ upon decoupling from the SM bath. When $\lambda_{\rm FS} \gtrsim k^{-1}_{\rm RH}$, gravitational lensing and pulsar timing array searches cannot put a bound on $T_{\rm RH}$ since ultra compact halo objects do not exist in the present universe~\cite{Erickcek:2011us}, as shown by the dashed orange line of Fig.~\ref{fig:signals}. From this bound, we see that probing reheat temperatures above $4 ~{\rm MeV} $ through observations of ultra compact DM halos requires $M_1 \gtrsim  ~{\rm MeV} $, which is already excluded by the insufficient dilution of $N_1$ DM. 

\subsection{Leptogenesis}
Besides providing an excellent DM candidate in the form of $N_1$, right-handed neutrinos are also appealing in that they can generate the observed baryon asymmetry via leptogenesis~\cite{Fukugita:1986hr}. In a forthcoming paper~\cite{future}, we show that the decay of a heavier, long-lived right-handed neutrino, $N_2$, can not only provide the dilution necessary to realize $N_1$ DM, but also generate a large lepton asymmetry. In the usual way, this lepton asymmetry is converted to a baryon asymmetry via electroweak sphalerons, generating the observed baryon asymmetry of our universe. Since the sphaleron process ceases operation at temperatures below the weak scale, baryogenesis is suppressed when $T_{\rm RH} < v$. In this case, the baryon asymmetry is generated by the fraction $(T_{\rm RH}/v)^2$ of $N_2$ that decay in the $N_2$ MD-era before the temperature of the universe falls below the weak scale.%
\footnote{When $v \simeq T_{\rm RH}$, the thermal bath is not primordial but generated by $N_2$ itself (see e.g.~\cite{Kolb:1990vq,Co:2015pka}), and the suppression is $(T_{\rm RH}/v)^4$.}
Consequently, the generated baryon asymmetry is
\begin{align}
	Y_B = \frac{28}{79} \times \epsilon \frac{3}{4}\frac{T_{\rm RH}}{M_2}\left( \frac{T_{\rm RH}}{v}\right)^2	 
	\label{eq:lepto}
\end{align}
where $\epsilon$ is the lepton asymmetry generated per $N_2$ decay, and the factor of $28/79$ accounts for the conversion of the lepton asymmetry into the baryon asymmetry via sphalerons~\cite{Harvey:1990qw}. 

Independent of the model, $\epsilon$ is at most unity.%
\footnote{Large $\epsilon$ requires a large Yukawa coupling $y_{33}$, which naively produces too large SM neutrino masses by the see-saw from $N_3$. This can be avoided by a certain structure in $y_{ij}’$ and $y_{ij}$. The magnitude of $\epsilon$ is also restricted by the stability of $N_1$ against quantum correction from $y_{33}$, further constraining the parameter space. We study this in detail in a future work.}
Conservatively taking this maximum $\epsilon$, we see from Eq.~(\ref{eq:lepto}) that generating the observed baryon asymmetry, $Y_B \simeq 8 \times 10^{-11}$, is impossible when $T_{\rm RH} \ll v$, as shown by the dashed purple contour of Fig.~\ref{fig:signals}. This constraint demonstrates that incorporating leptogenesis into $N_1$ DM from LR models severely diminishes the viable parameter space, and that future $21$-cm cosmological probes of warm DM can significantly probe this reduced parameter space.

\section{Predictions on $v_R$ from UV physics}

The cosmologically allowed region of initially thermalized $N_1$ DM in LR theories constrains the $SU(2)_R$ symmetry breaking scale $v_R$ well above the electroweak scale. As discussed in Sec.~\ref{sec:cosmology},
the viable region of the right-handed breaking scale is $ 10 ^{ 6} \lesssim v _R / {\rm GeV}  \lesssim 10 ^{ 13}  $, for any $c \leq 1$, and $ 10 ^{ 8} \lesssim v _R /  {\rm GeV} \lesssim 10 ^{ 13} $ for the case of $c=1$. In this section, we consider the implications such a breaking scale has on prospective theories behind LR models.

\subsection{Small Higgs quartic coupling at high energy scales}

Intriguingly, this range of $v_R$ is predicted independently within `Higgs-Parity' theories~\cite{Hall:2018let,Dunsky:2019api,Hall:2019qwx,Dunsky:2019upk}, a subset of LR models with Higgs-doublets $H_L$ and $H_R$ and with the LR symmetry spontaneously broken by $\vev{H_R} \gg \vev{H_L}$. In Higgs-Parity models, the SM Higgs quartic coupling $\lambda$  is predicted to vanish at the scale $v_R$. The SM renormalization group flow of $\lambda$ shows that $\lambda = 0$ for $ 10 ^{ 9}  \lesssim v _R /  {\rm GeV}\lesssim 10 ^{ 13}$, with an uncertainty dominantly arising from an uncertainty in the top quark mass~\cite{Buttazzo:2013uya}.

\begin{figure}[h]
	\begin{center}
	    \includegraphics[width=10.3cm]{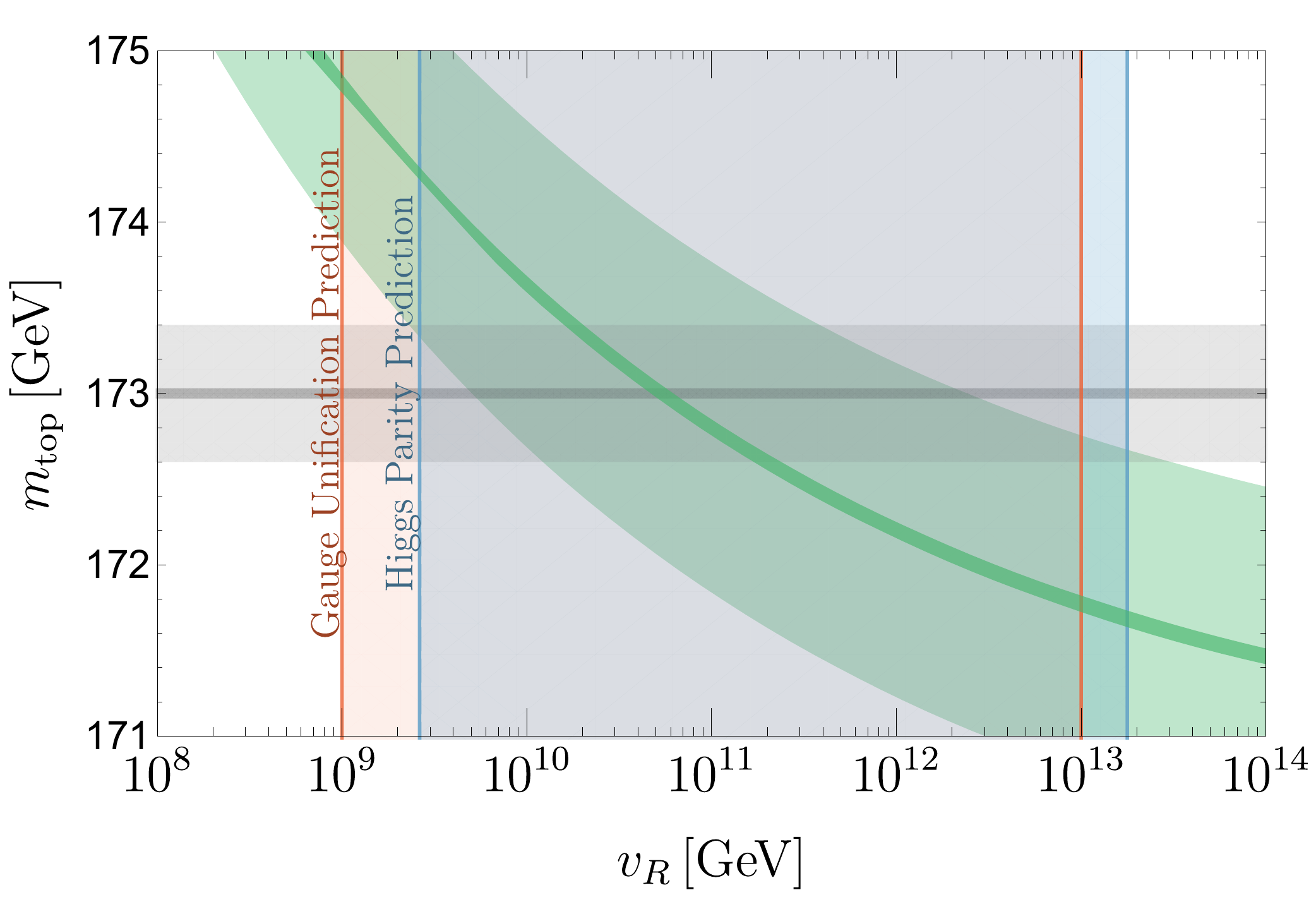}
	\end{center}
	\caption{The predicted top quark mass in Higgs Parity theories is shown in  {\bf \color{c3} green}, as a function of the right-handed symmetry breaking scale.  The experimentally preferred top mass is shown as a {\bf \color{gray}gray} band, leading to the preferred range of $v_R$ shown by the vertical {\bf \color{c1} blue} band.   The {\bf \color{c4} red} band shows the range of $v_R$ preferred by gauge coupling unification.}
	\label{fig:vRPrediction}	
\end{figure}

The is shown explicitly in Fig.~\ref{fig:vRPrediction}. The green band shows the relation between $v_R$ and the top quark mass, $m_{\rm top}$.%
\footnote{We ignore a UV completion-dependent part of the threshold correction to $\lambda(v_R)$ from $m_{\rm top}$ that in some extreme cases can lower the value of $v_R$ by $1-2$ orders of magnitude~\cite{Hall:2019qwx}.}
The width of this green band arises from the uncertainty of the Higgs mass $m_h = 125.18 \pm 0.16 $ GeV and the strong coupling constant $\alpha_s(m_Z) = 0.1181 \pm 0.0011$ at $2 \sigma$~\cite{Tanabashi:2018oca}.
The preferred value of the top quark mass (2$\sigma$) is shown by a horizontal gray band. As a result, the LR symmetry breaking scale $v_R$ is predicted to be $ 10 ^{ 9}  \lesssim v _R /  {\rm GeV}\lesssim 10 ^{ 13}$.
The narrower green band shows the relation assuming that the uncertainties shrink to $m_h = 125.18  \pm .020 $ GeV and  $\alpha_s(m_Z) = 0.1181 \pm 0.0001$, which is possible through improved lattice calculations, measurements at future lepton colliders, and measurements at HL-LHC~\cite{Cepeda:2019klc,Lepage:2014fla,Gomez-Ceballos:2013zzn}.
The top quark mass can be measured with an accuracy of a few tens of MeV by $e^+ e^-$ colliders~\cite{Seidel:2013sqa,Horiguchi:2013wra,Kiyo:2015ooa,Beneke:2015kwa} such as ILC~\cite{Baer:2013cma}, narrowing down the prediction on $v_R$ within a few tens of percent, as shown by the narrower gray band.
In future work, we will incorporate leptogenesis from $N_2$ decays with $N_1$ DM within the Higgs Parity framework~\cite{future}.

\subsection{Gauge coupling unification}
The cosmologically allowed range of $v_R$ is also consistent with gauge coupling unification.
The LR symmetric gauge group, $SU(3)_c\times SU(2)_L\times SU(2)_R \times U(1)_{B-L}$ is a subgroup of an $SO(10)$ unified gauge group. Assuming the minimal symmetry breaking chain containing the LR symmetric gauge group as an intermediate scale gauge group,
\begin{align}
SO(10) \longrightarrow
SU(3)_c\times SU(2)_L\times SU(2)_R\times U(1)_{B-L} \stackrel{v_R}{\longrightarrow} \;\;\; SU(3)_c\times SU(2)_L \times U(1)_Y,
\end{align}
the scale $v_R$ is predicted to be $ 10 ^{ 9}  \lesssim v _R /  {\rm GeV}\lesssim 10 ^{ 13}$~\cite{Rizzo:1981jr,Siringo:2012bc,Hall:2019qwx}.

We note, however, that a stable right-handed neutrino, $N_1$, is in tension with matter unification. In fact, if the SM quarks and leptons as well as the right-handed neutrinos are unified into a ${\bf 16}$ representation of $SO(10)$, Yukawa unification naively predicts that the right-handed neutrinos are all heavy and unstable. To evade this naive expectation would require a more sophisticated model in a four-dimensional $SO(10)$ unified theory. This could be possible with $SO(10)$ unification in higher dimensions with orbifolding~\cite{Kawamura:1999nj,Kawamura:2000ev,Altarelli:2001qj,Hall:2001pg}, where Yukawa couplings do not necessarily unify if matter is localized on gauge symmetry breaking branes~\cite{Hebecker:2001wq}. Even if matter lives in the bulk, the SM quarks and leptons as well as the right-handed neutrinos may arise from zero-modes of different ${\bf 16}$’s by the orbifold projections, as is realized in $SU(5)$~\cite{Hall:2001pg,Hebecker:2001wq} or $SO(10)$~\cite{Hall:2001xr} unification without intermediate gauge symmetry. Breaking of $SO(10)$ down into LR symmetry by orbifolding is discussed in~\cite{Biermann:2019amx}.

\section{Freeze-In}

When the reheat temperature of the universe is below the thermalization temperature of the right-handed neutrinos (see~\eqref{eq:eqtemp}), neither $N_1$ nor $ N _2 $ have a thermal abundance. Instead, the $ N _1 $ abundance is determined by scattering via heavy $W_R$ and $Z_R$ exchange which, being UV-dominated, depends on the reheating temperature,
\begin{align}
\frac{\rho_{N_1}}{s} & \simeq 1 \times 10 ^{ - 5 }\frac{M _1 \left(T^{\rm inf} _{ {\rm RH}}\right) ^3 M_{\rm pl}  }{ v _R ^4 } \,,\\ 
\Rightarrow \frac{ \Omega }{ \Omega _{ {\rm DM}}} & \simeq \left(\frac{M_1}{150 \, \KEV}\right)\left(\frac{10^{10} ~\GEV}{v_R}\right)^4  \left(\frac{T^{\rm inf}_{\rm RH}}{10^{7} ~\GEV}\right)^3\,.
\label{eq:FIAbundance}
\end{align}
Freeze-in production from other sources, such as $ \ell H \rightarrow N _1  $, are subdominant since $ y _{ 1i} \ll 1 $ is needed to ensure $ N _1 $ is long-lived. 
Contributions to the $N_1$ abundance may also arise from beta decays of $N_2$ and $N_3$. These, however, are always subdominant to the direct freeze-in production of $N_1$, whether $N_{2,3}$ are produced by the $W_R$ interaction or the $\ell N H$ interaction. 

\begin{figure}[h]
	\begin{center}\includegraphics[width=0.8\linewidth]{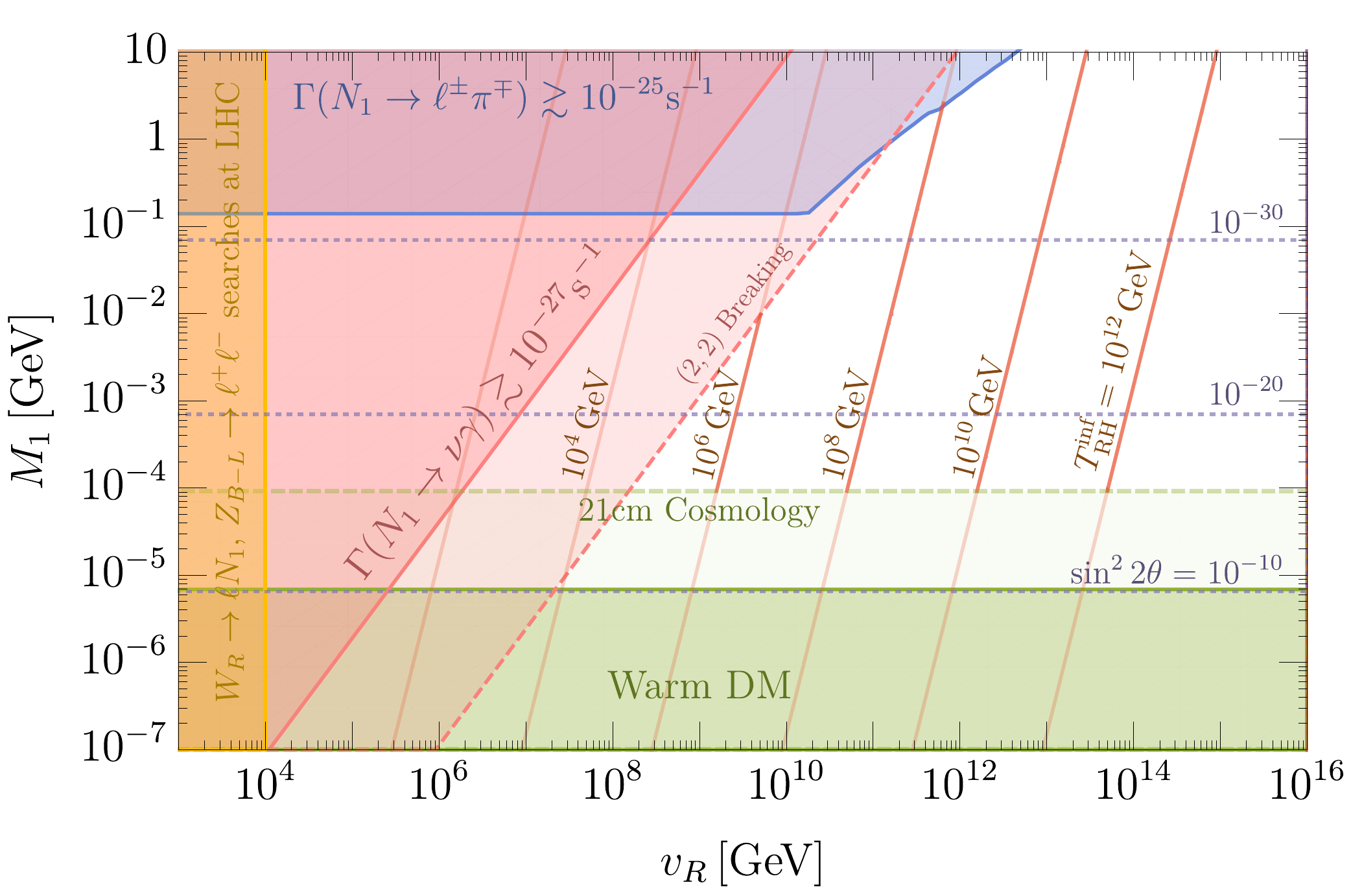}\end{center}
	\caption{The parameter space for $N_1$ DM produced by freeze-in. The observed relic abundance occurs in the unshaded region for values of $T^{\rm inf}_{\rm{RH}}$ shown by the dashed {\color{c4} \bf red} contours. Constraints from small scale structure are shown in {\color{c3} \bf green}, with projections from future probes of small scale structure using the 21cm line in dashed {\color{c3} \bf green}.  In the {\color{c1} \bf blue} region $N_1$ decays too rapidly via $W_R$ to $\ell^\pm \pi^\mp$, and in the {\color{pink} \bf pink} region $N_1$ decays too rapidly via $W_R-W_L$ mixing to  $\nu \gamma$ when $SU(2)_L$ is broken by $(2,1)+(1,2)$ (solid) or by $(2,2)$ (dashed). The decay via $W_R-W_L$ mixing to $\ell^+ \ell^- \nu$ is weaker and not shown. The horizontal dashed {\color{c1} \bf  blue} lines show the limit (\ref{eq:sin2theta}) on the mixing angle of $N_1$ with active neutrinos. Collider searches for $W_R$ exclude $v_R$ below about $10 \, \TEV$, as shown in {\color{c2} \bf orange}.}
	\label{fig:thermal_FI}	
\end{figure}

In Fig.~\ref{fig:thermal_FI}, we show the contours of the required reheat temperature after inflation to freeze-in $N_1$ DM for a given $(v_R,M_1)$. In the green region, the warmness of $N_1$ affects large scale structure. Since frozen-in $N_1$ is never diluted, it is warmer than frozen-out $N_1$ for a fixed $M_1$. More concretely, its free-streaming length is larger by a factor of approximately $\frac{4}{3.2}\left(\frac{M_1 Y_{\rm therm}}{\rho_{\rm DM}/s}\right)^{1/3}$, which gives a commensurately stronger warm DM bound compared to Fig.~\ref{fig:signals}. Here, the factor of $4/3.2$ comes from the difference in $\langle p/T \rangle$ for the non-thermal frozen-in distribution, to the thermal frozen-out distribution, as discussed in~\cite{Heeck:2017xbu}. In the blue and pink regions, the decay of $N_1$ mediated by $W_R$, \eqref{eq:WRdecay}, or $W_R-W_L$ mixing, \eqref{eq:N1WRWL2}, overproduces the observed amount of galactic gamma-rays, respectively~\cite{Essig:2013goa}. Similarly, the decay of $N_1$ via active-sterile mixing overproduces the observed galactic x-rays and gamma-rays for the mixing angle $\sin^2 2\theta_1$ labeling the purple dotted contours. Unlike the $W_R$-mediated decay, which is fixed by $v_R$, the decay via $N_1-\nu$ mixing is set by the free parameter $\theta_1$. Lastly, searches at the LHC for heavy charged boson resonances ($pp \rightarrow W_R \rightarrow N_1 \ell)$~\cite{Aad:2019wvl} and neutral boson resonances ($pp \rightarrow Z_R \rightarrow \ell^+ \ell^-)$~\cite{Aad:2019fac} exclude $v_R$ below about $10 \,\TEV$, as shown by the orange region.

Fig.~\ref{fig:thermal_FI} shows that the parameter space for $N_1$ DM from freeze-in is weakly constrained compared to that of $N_1$ DM from freeze-out and dilution, shown in Fig.~\ref{fig:thermal_FO}. For example, $v_R$ could be as low as about 100 TeV, with the reheat temperature after inflation below $100 \, \GEV$. Likewise, bounds on $M_1$ are weak; although as $M_1$ increases $\sin^2 2\theta_1$ is constrained to become extremely small to keep $N_1$ sufficiently long-lived. However, if leptogenesis via $N_2$ decay is incorporated into the $N_1$ DM freeze-in cosmology, the $(M_1,v_R)$ parameter space becomes more tightly constrained. In a future work, we discuss this viable parameter space in the framework of Higgs Parity~\cite{future}.

\section{Conclusions}
Since right-handed neutrinos $N_i$ have no SM gauge interactions, it is plausible that one of them, $N_1$, is sufficiently stable to make up dark matter. A theory containing $N_i$ has three types of neutrino masses: Dirac masses, $(\nu_i N_j)$, and Majorana masses, $(\nu_i \nu_j)$ and $(N_i N_j)$. In general, these are described by three independent mass matrices.  In this paper we have studied theories with a LR symmetry that forces the Majorana mass matrices for $(\nu_i \nu_j)$ and $(N_i N_j)$ to be proportional.  In simple theories with the $SU(2)_R$ and $SU(2)_L$ gauge groups broken by doublet vacuum expectation values of strength $v_R$ and $v$, the constant of proportionality is $v^2/v_R^2$, whereas in the conventional LR theory, with scalar triplets and bidoublets, the constant of proportionality is $c \, v^2/v_R^2$, with $c \lesssim 1$.

At sufficiently high temperatures in the early universe, $N_i$ are kept in thermal equilibrium via the $SU(2)_R$ gauge interactions. The initially thermal $N_1$ can account for the observed DM if they are subsequently diluted by decays of the initially thermal $N_2$. We have shown that the $(v_R, M_1)$ parameter space for this simple origin of DM is highly restricted, and indeed bounded, as shown in Fig.~\ref{fig:thermal_FO}. The allowed region is triangular with $v_R \simeq (10^8 - 3 \times 10^{12})$ GeV and $M_1 \simeq (2 \, \KEV - 1 \, \MEV)$ for $c=1$. As $c$ is reduced, the allowed region shrinks in size and shifts to lower values of $(v_R, M_1)$, disappearing entirely at $(10^6 \, \GEV,2 \, \KEV)$ when $c \simeq 10^{-4}$.

Constraints that determine the lower bounds on $M_1$ and $v_R$ are straightforward, arising from requirements that DM not be too warm and $N_2$ decays without disturbing nucleosynthesis. However, there is a third constraint, which leads to upper bounds on both $M_1$ and $v_R$, and is involved.  In Appendix~\ref{sec:nuMasses} we show that this DM scenario places constraints on the active neutrino masses in such a way that the mass of $N_2$ is determined by (\ref{eq:M2m2}), and grows rapidly with $v_R$.  Thus at large enough $v_R$, $N_2$ decays dominantly via $W_R$ exchange; the requirement that this decay is slow enough to sufficiently dilute $N_1$ places an upper bound on $M_1 v_R$, as shown by the blue region of Fig.~\ref{fig:thermal_FO}.  

Observational probes of this $N_1$ DM, from relativistic freeze-out and dilution by $N_2$ decay, are shown in Fig.~\ref{fig:signals} for $c=1$. The bulk of the $(v_R, M_1)$ parameter space is at lower values of $M_1$, leading to signals of warmness in large scale structure.  Indeed, a significant portion of the parameter space can be observationally probed using 21 cm cosmology. A subdominant component of $N_1$ DM is produced non-thermally via the $W_R$ beta decay $N_2 \rightarrow N_1 \ell^+ \ell^-$, producing $N_1$ that become non-relativistic at temperatures ${\cal O} ({\rm eV})$ and are therefore hot. The size of this component is proportional to $(M_1 v_R)^2$ and, coincidentally, present limits on this hot DM component are close to the previously described limit on $M_1 v_R$ from too much $N_1$ DM. Indeed, the interesting case of $N_2$ decaying dominantly via $W_R$ is already in tension with observation, and future CMB measurements will thoroughly probe this possibility.   During the era of $N_2$ matter domination, density perturbations on small enough scales grow and could potentially lead to observable structures.  Unfortunately, for pulsar timing arrays to see a signal in the region of reheat temperatures above the 4 MeV BBN bound, requires $M_1 > \MEV$,  which is excluded by insufficient dilution of $N_1$. 

Given that the decays of $N_2$ are out of thermal equilibrium, it is plausible that they lead to leptogenesis.  We explore this is detail in a future publication~\cite{future}, and here we simply observe that sufficient baryon asymmetry arises only if such decays are early enough, as shown by the dashed purple line in Fig.~\ref{fig:signals}. A large fraction of the parameter space that allows leptogenesis can be probed by 21 cm cosmology.

The $SU(2) \times SU(2)_R \times U(1)_{B-L}$ gauge group studied in this paper provides an elegant setting for Higgs Parity~\cite{Hall:2018let,Dunsky:2019api,Hall:2019qwx,Dunsky:2019upk}, which correlates the SM parameters including the top quark and Higgs boson masses and the QCD coupling constant with the scale of $SU(2)_R$ breaking.  The predicted top quark mass in this scheme is consistent with the experimentally preferred value of it for $v_R$ in the range of $(10^9 - 10^{13})$ GeV, as shown in Fig.~\ref{fig:vRPrediction}, which includes much of the range relevant for $N_1$ DM. As uncertainties in the Higgs mass and the QCD coupling are reduced in near future measurements, $v_R$ is predicted within a factor of 10. It will be interesting to see whether the ranges of $v_R$ for Higgs Parity and $N_1$ DM remain consistent. Precise measurements of the top quark mass at future linear colliders such as ILC can predict $v_R$ with an accuracy of a few tens of percent.   The range of $v_R$ that gives precision gauge coupling unification is also shown in Fig.~\ref{fig:vRPrediction}; remarkably it is consistent with Higgs Parity and much of the range needed for $N_1$ DM.  An important question is how easily the conditions for cosmological stability of $N_1$ can be implemented in a realistic $SO(10)$ theory of flavor.

In LR theories, if the reheat temperature after inflation is too low for $W_R$ exchange to put $N_i$ into thermal equilibrium, the $N_1$ DM abundance can be successfully generated by freeze-in, as shown by the solid red contours in Fig.~\ref{fig:thermal_FI}. In this case the scale $v_R$ is unconstrained, except by direct limits from LHC on the masses of $W_R$ and $Z_R$.  There are, however, strong limits on $M_1$ from warmness and from $N_1$ stability requirements.

\section*{Acknowledgement}
We thank Bibhushan Shakya for useful discussion. This work was supported in part by the Director, Office of Science, Office of High Energy and Nuclear Physics, of the US Department of Energy under Contracts DE-AC02-05CH11231 (JD and LJH) and DE-SC0009988 (KH), as well as by the National Science Foundation under grants PHY-1316783 and PHY-1521446 (LJH).

\appendix

\section{Neutrino mass relations}
\label{sec:nuMasses}
In this appendix, we show the constraints on the mass eigenvalues of the active neutrinos through the requirements of abundance and radiative stability of $N_1$ DM, together with cosmological bounds on the warmness of $N_1$ and the reheating temperature from $N_2$ decay.  We remind the reader that we work in a mass basis for $N_i$, which have masses $M_i$.  The states $\nu_i$ are related to $N_i$ by LR symmetry, and are not necessarily mass eigenstates.

We first consider the case $M_3 > M_1$. 
Constraints on the Yukawa matrix $y_{ij}$, and lower bounds on $M_1$ and $M_2$, then ensure that the seesaw mechanism is operative, so that the $\nu_i$ mass matrix is
\begin{align}
	m_{ij} &= \delta_{ij}c \frac{v^2}{v_R^2} M_i - \sum_{k=1}^3 \frac{y_{ik} y_{jk}}{M_k} v^2
	\label{eq:nuMassMatrixApp}
\end{align}
as in (\ref{eq:numassmatrix}). We will demonstrate two claims:
\begin{enumerate}[label={Claim \arabic*:},wide = 0pt, font =\bf]
\item The lightest eigenstate is aligned with $\nu_1$, with mass  $m_1 \ll \sqrt{\smash[b]{\Delta m^2_{\rm sol}}} \simeq 0.01 \, \EV$.
\item The mass of $N_2$ is determined by $M_2 \simeq \mu (v_R/v)^2 c^{-1}$, \rm{where} $0.01 \, \EV \lesssim \mu \lesssim 0.10 \, \EV$. This is key to constraining the parameter space of frozen-out $N_1$ DM. 
\end{enumerate}

The stability of $N_1$ and $N_2$ require $|y_{1i}|, |y_{2i}| \ll 1$, as indicated by Eqs.~\eqref{eq:yi1} and \eqref{eq:yi2}, implying that the seesaw contributions from $N_1$ and $N_2$ exchange are both much less than 0.01 eV.  Hence, to an excellent approximation, Eq.~\eqref{eq:nuMassMatrixApp} can be written as
\begin{align}
\label{eq:numass1}
m_{ij} \simeq
\begingroup 
\setlength\arraycolsep{5pt}
\renewcommand{\arraystretch}{3}
\begin{pmatrix}
	c \left(\dfrac{v}{v_R}\right)^2 M_1 - \dfrac{y_{13}^2}{M_3}v^2 & - \dfrac{y_{23}y_{13}}{M_3}v^2 & - \dfrac{y_{13}y_{33}}{M_3}v^2 \\
	- \dfrac{y_{23}y_{13}}{M_3} & c\left(\dfrac{v}{v_R}\right)^2 M_2 - \dfrac{y_{23}^2}{M_3}v^2 & - \dfrac{y_{23}y_{33}}{M_3}v^2 \\
	- \dfrac{y_{13}y_{33}}{M_3}v^2 & - \dfrac{y_{23}y_{33}}{M_3}v^2 & c \left(\dfrac{v}{v_R}\right)^2 M_3 - \dfrac{y_{33}^2}{M_3}v^2
\end{pmatrix}
\endgroup .
\end{align}
Next we find that the entry $m_{11}$ is much smaller than $\sqrt{\smash[b]{\Delta m^2_{\rm sol}}}$:
\begin{align}
	c\left(\dfrac{v}{v_R}\right)^2 M_1 &\leq v^2 M_1\left(\frac{1536\pi^3}{14 M_2^5 M_{\rm Pl}} \left(\frac{\pi^2 g_*(T_{\rm RH})}{10}\right)^{1/2} \left(\frac{\rho_{\rm DM}/s \, M_2}{1.6 \frac{3}{4} M_1}\right)^2\right)^{1/2} \tag{$N_2$ stability} \nonumber \\
	&= 6 \times 10^{-6} \, \EV \left(\frac{24 \, \GEV}{M_2}\right)^{3/2} \left(\frac{g_*(T_{\rm RH})}{10.9}\right)^{1/4},\\
\dfrac{|y_{13}|^2}{M_3} v^2 &\leq \frac{M_1^2}{M_3}\sin^2 2\theta_1 \nonumber \\ \tag{$N_1$ stability}
 & \leq 8 \times 10^{-5} \, \EV \left(\frac{2 \, \KEV}{M_1}\right)^{4} \left(\frac{M_1/M_3}{1}\right).
\end{align}
Now we argue that $m_{13}$ is also negligible. The upper bound on $|y_{13}|$ of \eqref{eq:yi1} from the stability of $N_1$ implies that $m_{13}$ is non-negligible only if $|y_{33}|$ is large, such that $|y_{33}|^2 v^2/M_3 \gg \sqrt{\smash[b]{\Delta m^2_{\rm sol}}}$. To ensure that the observed sum of neutrino masses does not exceed $0.06-0.10 \, \EV$, $m_{33}$ must be tuned such that $ |y_{33}|^2 v^2/M_3 \simeq c(v/v_R)^2 M_3$. However,
\begin{align}
		\dfrac{|y_{13}y_{33}|}{M_3} v^2 &\simeq \sqrt{c}\dfrac{|y_{31}| v^2}{v_R}  \\
		&\leq \sqrt{c}M_1 \sin \theta_1 \dfrac{v}{v_R}  \nonumber \\
		&\leq M_1 \sin \theta_1 v \left(\frac{1536\pi^3}{14 M_2^5 M_{\rm Pl}} \left(\frac{\pi^2 g_*(T_{\rm RH})}{10}\right)^{1/2} \left(\frac{\rho_{\rm DM}/s \, M_2}{1.6 \frac{3}{4} M_1}\right)^2\right)^{1/4} \tag{$N_2$ stability} \nonumber \\
		 &\leq 2 \times 10^{-5}\, \EV \left(\frac{M_1}{2 \,\KEV} \right)^{-2} \left(\frac{M_2}{24 \,\GEV} \right)^{-3/4} \tag{$N_1$ stability}. \nonumber
\end{align}
Hence, from the lower bounds on $M_{1,2}$ we conclude that $m_{13}$ is negligible. 

The mass matrix of the active neutrinos is therefore approximately
\begin{align}
\label{eq:numass2}
m_{ij} \simeq
\begingroup 
\setlength\arraycolsep{5pt}
\renewcommand{\arraystretch}{3}
\begin{pmatrix}
	0 & - \dfrac{y_{23}y_{13}}{M_3}v^2 & 0 \\
	- \dfrac{y_{23}y_{13}}{M_3} & c\left(\dfrac{v}{v_R}\right)^2 M_2 - \dfrac{y_{23}^2}{M_3}v^2 & - \dfrac{y_{23}y_{33}}{M_3}v^2 \\
	0 & - \dfrac{y_{23}y_{33}}{M_3}v^2 & c \left(\dfrac{v}{v_R}\right)^2 M_3 - \dfrac{y_{33}^2}{M_3}v^2
\end{pmatrix}
\endgroup.
\end{align}
We put further constraints on the mass matrix by considering the two cases of $M_3$: greater than or less than $M_2$.

\subsection*{Case 1:  $M_3 > M_2$}

For this case, the entry $m_{12}$ is negligible. This is
because the upper bound on $y_{32}$ is 
\begin{align}
|y_{23}|^2 &\leq \frac{1}{\Gamma_0 M_{\rm Pl}} \left(\frac{\pi^2 g_*(T_{\rm RH})}{10}\right)^{1/2} \left(\frac{ M_2\,\rho_{\rm DM}/s }{1.6 \frac{3}{4} M_1}\right)^2, \tag{$N_2$ stability}   \\
\Gamma_0 &\equiv \begin{cases}
\dfrac{171/8}{1536 \pi^3}\dfrac{M_2^3}{v^2} & M_2 < v \vspace{0.5em} \\ 
\dfrac{1}{8\pi}M_2 & M_2 > v,  \\ 
\end{cases}
\end{align}
so that
\begin{align}
|m_{12}| &= \frac{|y_{13} y_{23}| v^2 }{M_3} \\
 &\leq  \sin \theta_1 v \left(\frac{1}{\Gamma_0 M_{\rm Pl}} \left(\frac{\pi^2 g_*(T_{\rm RH})}{10}\right)^{1/2} \left(\frac{\rho_{\rm DM}/s}{1.6 \frac{3}{4}}\right)^2\right)^{1/2}.  \tag{Stability of $N_1$ and $N_2$, $M_3 > M_2$} \\
 &\leq 9 \times 10^{-10} \, \EV \left(\frac{M_1}{2 \, \KEV} \right)^{-5/2} \left(\frac{M_2}{24 \, \GEV} \right)^{-3/2} 
\end{align}
Next we show that $m_{23}$ is also small. The upper bound on $y_{33}$ is
\begin{align}
|y_{33}|^2 &=  \frac{M_3^2}{v^2}\left|\frac{m_{22} + \frac{y_{23}^2}{M_3}v^2}{M_2} - \frac{m_{33}}{M_3}\right| \tag{Rewriting $m_{33}$} \\
                 & \leq  \frac{M_3^2}{v^2} \left( \left| \frac{m_{22}}{M_2}\right| +\left| \frac{y_{23}^2v^2}{M_2 M_3} \right| +\left| \frac{m_{33}}{M_3} \right|  \right) \tag{Triangle inequality}  \\
		&\leq \frac{M_3^2}{v^2 M_2}\left(  \left|m_{22}\right| + \left| \frac{y_{23}^2 v^2 }{M_2}\right| + \left| m_{33} \right|  \right) \tag{$M_2 < M_3$}  \\
		&\lesssim \frac{M_3^2}{v^2}\frac{\sum m_i }{M_2}.  \tag{Upper bound on $m_{22}$ and $m_{33}$, $N_2$ stability} 
\end{align}
Hence, $m_{23}$ is at most
\begin{align}
|m_{23}| &\leq v \sqrt{\left(\frac{1}{\Gamma_0 M_{\rm Pl}} \left(\frac{\pi^2 g_*(T_{\rm RH})}{10}\right)^{1/2} \left(\frac{M_2\,\rho_{\rm DM}/s}{1.6 \frac{3}{4} M_1}\right)^2 \right) \left(\frac{\sum m_i}{M_2} \right)}. \label{eq:m32bound}
\end{align}
Fig.~\ref{fig:m2case12}$\,$({\bf left}) shows the region where the right-side of Eq.~\eqref{eq:m32bound} is greater than $\sqrt{ \smash[b]{ \Delta m_{\rm sol}  ^2}}$ in the $M_1-M_2$ plane. As can be seen, everywhere in the cosmologically allowed region $|m_{23}| \ll \sqrt{ \smash[b]{ \Delta m_{\rm sol}  ^2}}$. In the active neutrino mass matrix, only $m_{22}$ and $m_{33}$ can be comparable to the observed neutrino masses; for $M_3 > M_2$ the $\nu_i$ basis is accurately the mass basis.
The lightest active neutrino mass $m_1$ is much smaller than $\sqrt{\smash[b]{\Delta m^2_{\rm sol}}}$, showing {\bf Claim 1}.

The two heavier active neutrino masses ($m_2, m_3$) are simply given by
\begin{align}
	\label{eq:m2case1}
 	m_{2} &\simeq m_{22} =  c \left(\frac{v}{v_R}\right)^2 M_2 - \frac{y_{23}^2}{M_3}v^2 \\
 	m_{3} &\simeq m_{33} =  c \left(\frac{v}{v_R}\right)^2 M_3 - \frac{y_{33}^2}{M_3}v^2
\end{align} 
Furthermore,
\begin{align*}
\frac{|y_{23}|^2}{M_3}v^2 &\leq \frac{|y_{23}|^2}{M_2}v^2 \tag{$M_2 < M_3$} \\
						&\ll \sqrt{ \smash[b]{ \Delta m^2_{\rm sol}}}   \tag{$N_2$ stability}.
\end{align*}
Therefore, we obtain {\bf Claim 2}, with $\mu$ identified as $m_2$, the mass of $\nu_2$
\begin{align}
M_2 &\simeq m_2 \left(\frac{v_R}{v}\right)^2 \frac{1}{c}.
\end{align}

\subsection*{Case 2: $M_3 < M_2$}

We first show that $|y_{23}|^2 v^2 / M_3$ cannot be larger than the active neutrino mass by contradiction.
Let us assume that $|y_{23}|^2 v^2 / M_3 $ is larger than the active neutrino mass. Then to suppress $m_{22}$, we need
\begin{align}
\frac{|y_{23}|^2}{M_3}v^2 &\simeq c\left(\frac{v}{v_R}\right)^2 M_2.
\end{align}
If $|y_{33}|$ is larger than $|y_{23}|$, $|y_{33}|^2 v^2 / M_3 $ is also larger than the active neutrino mass and must be cancelled by $c M_3 (v/v_R)^2$, which is impossible since $M_3 < M_2$. We conclude that  $|y_{33}| < |y_{22}|$, which is used later.

Since $M_3 >M_1$, the case where $N_3$ decays after matter-radiation equality is excluded due to entropy production by the decay, or too much $N_3$ DM if $N_3$ is cosmologically stable. We thus assume that $N_3$ decays before matter-radiation equality.

\subsubsection*{Case 2-1: $M_2 < v$}
Since $|y_{33}|< |y_{23}|$ and $|y_{13}|$ is small, the decay of $N_3$ by $W_L$ exchange is determined by $y_{23}$. Then the decay rate of $N_3$ by $W_R$ exchange is negligible. In fact, if $N_3$ decays dominantly by $W_R$ exchange,
\begin{align}
|y_{23}|^2 \frac{M_3^3}{v^2} < \frac{M_3^5}{v_R^4}.
\end{align}
In this case, however,
\begin{align}
\frac{|y_{23}|^2 v^2}{M_3} &\leq M_2 v^4\left(\frac{1536\pi^3}{14 M_2^5 M_{\rm Pl}} \left(\frac{\pi^2 g_*(T_{\rm RH})}{10}\right)^{1/2} \left(\frac{\rho_{\rm DM}/s \, M_2}{1.6 \frac{3}{4} M_1}\right)^2\right)\tag{$N_2$ stability, $M_2 > M_3$} \\
&= 2 \times 10^{-7} \, \EV \left(\frac{M_2}{24 \,\GEV}\right)^{-2}\left(\frac{M_1}{2 \,\KEV}\right)^{-2}
\end{align}
which is in contradiction. Thus $N_3$ decays dominantly by $y_{32}$.

In order for $N_2$ to be the diluter (by definition), it must be that
\begin{align}
\frac{M_2}{\sqrt{\Gamma_{N_2}}} &> \frac{M_3}{\sqrt{\Gamma_{N_3}}}, \tag{Dilution factor}\\
\frac{M_2}{\sqrt{|y_{23}|^2 M_2^3}} &> \frac{M_3}{\sqrt{|y_{23}|^2 M_3^3}} \tag{$\Gamma_{W_L} \propto y^2 M^3$} \\
\Rightarrow M_3 &> M_2,
\end{align}
which is a contradiction with our assumption that $M_3 < M_2$.

\subsubsection*{Case 2-2: $M_2 > v$}
When $M_2 > v$, $N_2$ decays to $\ell H$ via $y_{2i}$ or beta-decays via $W_R$ exchange. Both decay channels limit $|y_{23}|^2 v^2/M_3$ to ensure $N_2$ is long-lived enough to provide dilution of $N_1$.

\begin{figure}[]
    \centering
    \begin{minipage}[l]{0.5\textwidth}
        \centering
        \includegraphics[width=1.0\textwidth]{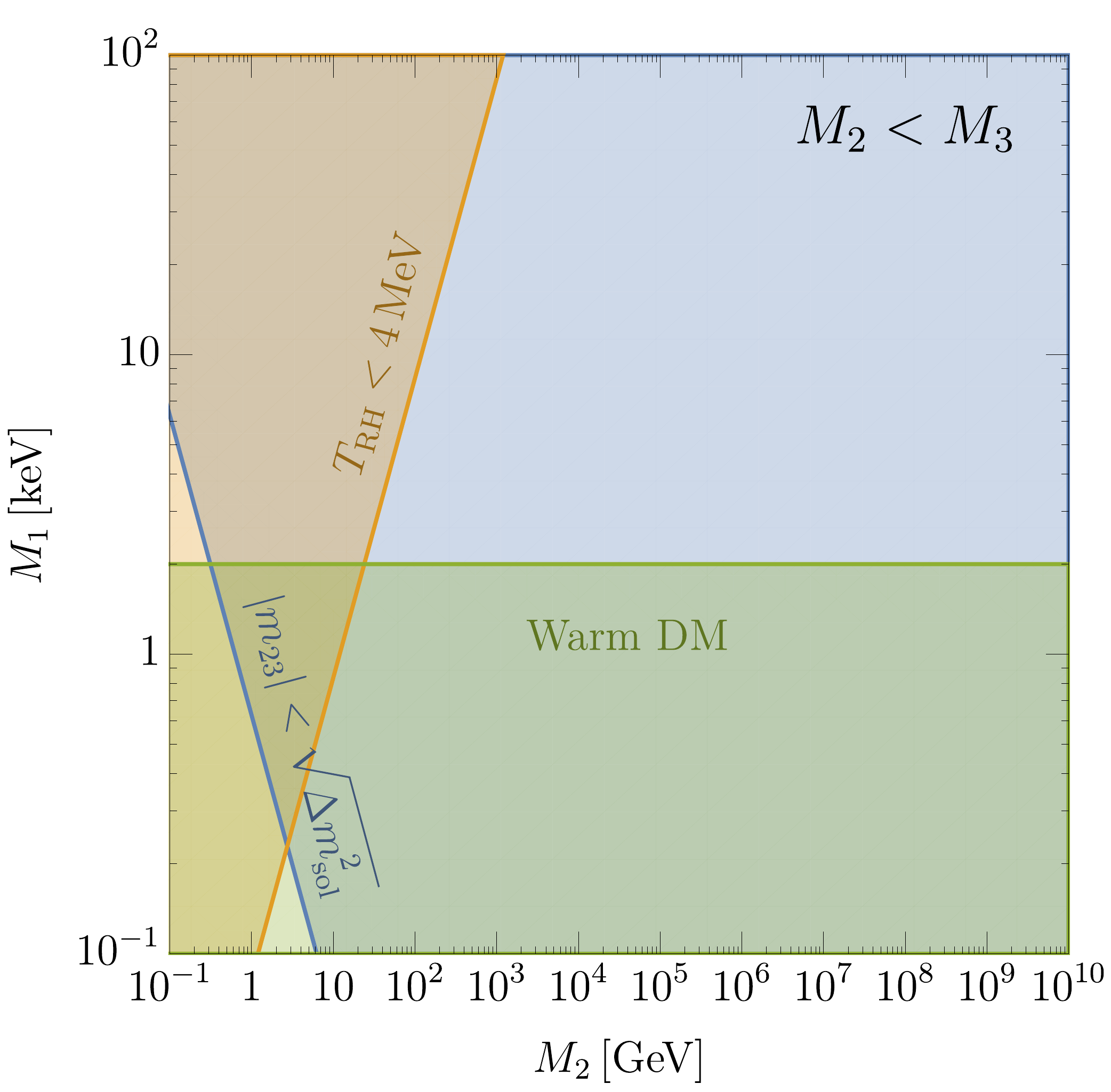}
    \end{minipage}\hfill
    \begin{minipage}[l]{0.5\textwidth}
        \centering
        \includegraphics[width=1.0\textwidth]{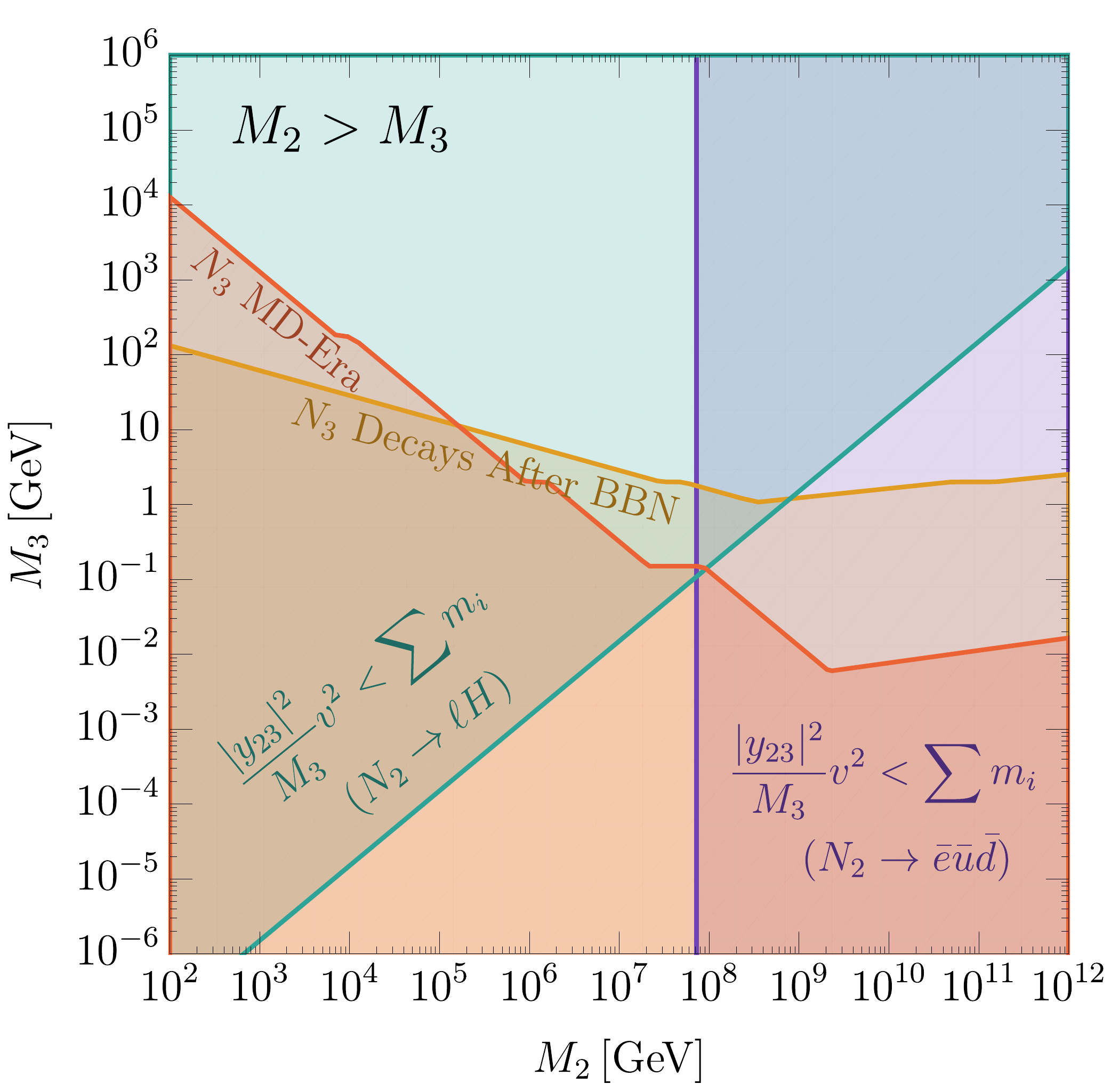} 
    \end{minipage}\hfill
    \caption{The right-handed neutrino mass parameter space showing the constraints which prove  {\bf Claim 1} and  {\bf Claim 2}. {\bf Left}: $M_2 < M_3$ (Case 1) -- the relation $M_2 = m_2 (v_R/v)^2/ c$ is guaranteed if $|m_{23}| < \sqrt{\smash[b]{\Delta m_{\rm sol}}^2}$. Stability of $N_2$ ensures $|m_{23}| < \sqrt{\smash[b]{\Delta m_{\rm sol}^2}}$ in {\bf \color{c1} blue}, which encompasses all of the parameter space not excluded by the warmness of DM ({\color{c3}\bf green}) or Big Bang Nucleosynthesis ({\bf \color{c2}  orange}). {\bf Right}: $M_2>M_3$ (Case 2) -- the relation $M_2 = \mu (v_R/v)^2/c$, where $0.01 \, \EV \lesssim \mu \lesssim 0.10 \, \EV$, is guaranteed if $|y_{23}|^2 v^2/M_3 < \sum m_i$. Stability of $N_2$ ensures $|y_{23}|^2 v^2/M_3 < \sum m_i$ in the {\bf \color{c5} purple} and {\color{turq}\bf turqoise} regions which encompass all of the parameter space not excluded from $N_3$ decaying after Big Bang Nucleosynthesis ({\bf \color{c2}  orange}). $N_3$ disrupts Big Bang Nucleosynthesis from the energy released in its decays when $M_3 > 1 \, \MEV$ in the {\bf \color{c2}  orange} region, and from the entropy produced from its decays in the intersection of the {\bf \color{c2}  orange} and  {\bf \color{c4}  red} regions.}
     \label{fig:m2case12}
\end{figure}

From the decay via $y_{2i}$,
\begin{align}
 \frac{|y_{23}|^2}{M_3}v^2 &\leq \frac{v^2}{M_3}\frac{8\pi}{M_2 M_{\rm Pl}} \left(\frac{\pi^2 g_*(T_{\rm RH})}{10}\right)^{1/2} \left(\frac{\rho_{\rm DM}/s \, M_2}{1.6 \frac{3}{4} M_1}\right)^2 \tag{$N_2$ stability}   
\end{align}
To be compatible with our assumption that $|y_{23}|^2 v^2/M_3 > m_1 + m_2 + m_3$, it is required that 
\begin{align}
 \frac{M_3}{M_2} &< \frac{8\pi v^2}{M_{\rm Pl} \sum m_i } \left(\frac{\pi^2 g_*(T_{\rm RH})}{10}\right)^{1/2} \left(\frac{\rho_{\rm DM}/s}{1.6 \frac{3}{4} M_1}\right)^2 \\
 			  &= 2 \times 10^{-9} \left(\frac{\sqrt{\smash[b]{ \Delta m_{\rm atm}  ^2}}}{\sum m_i} \right) \left(\frac{g_*({T_{\rm RH}})}{106.5} \right)^{1/2} \left(\frac{2 \, \KEV}{M_1} \right)^2.
\label{eq:m3m2limit}
\end{align}
The turquoise shaded region in Fig.~\ref{fig:m2case12}$\,$({\bf right}) violates this condition for the minimum value of $M_1 = 2 \, \KEV$; larger $M_1$ enlarges the region.
From the decay via $W_R$ exchange,
\begin{align}
\frac{|y_{23}|^2}{M_3}v^2 &\leq M_2 \left(\frac{v}{v_R}\right)^2 \tag{Since $c \leq 1$} \\
 				&\leq v^2\left(\frac{1536\pi^3}{20 M_2^3 M_{\rm Pl}} \left(\frac{\pi^2 g_*(T_{\rm RH})}{10}\right)^{1/2} \left(\frac{\rho_{\rm DM}/s \, M_2}{1.6 \frac{3}{4} M_1}\right)^2\right)^{1/2}. \tag{$N_2$ stability}  
\end{align}
In the purple region of Fig.~\ref{fig:m2case12}$\,$({\bf right}), the inequality is less than $\sum m_i$, also for the minimum value of $M_1 = 2 \, \KEV$.

There are additional constraints on $M_2$ and $M_3$ if $N_3$ decays after BBN.  This occurs when
\begin{align}
    \Gamma_{N_3} \simeq (2-20)\times \frac{1}{1536\pi^3}\frac{M_3^3}{v^2}|y_{23}^2| < (0.1 {\, \rm sec})^{-1},\,
\end{align}
where the coefficient depends on the kinematically available final states. If $M_3 >$ few MeV, then the decay products of $N_3$ carry enough energy to dissociate light elements formed during BBN, altering their relic abundances (see~\cite{Kawasaki:2004qu} and references therein).%
\footnote{For $M_3< 100$ MeV hadronic decays of $N_3$ are absent and the effect on BBN only comes from photo-dissociation which is effective for $T<0.01$ MeV. We find that $N_3$ decays below $T=0.01$~MeV for $M_3 < 100$~MeV.}
If $M_3 \lsim $ MeV, the decay after BBN does not necessarily dissociate any light elements, but can still alter their relic abundance if $N_3$ is long-lived enough to induce a matter-dominated era before decaying. This occur when 
\begin{align}
    \Gamma_{N_3} \lsim \left(\frac{\pi^2 g_*}{10}\right)^{1/2}\frac{1}{M_{\rm Pl}} \left(\frac{M_3}{M_1} \; \frac{\rho_{\rm DM}}{s} \right)^2\,.
\end{align}
These constraints are shown as the orange region of Fig.~\ref{fig:m2case12}$\,$({\bf right}), where we use the upper bound on $|y_{23}|$ from the stability of $N_2$ as discussed above and use the decay rate for $M_3 \gtrsim 2 m_e$. For smaller $M_3$, the actual decay rate is smaller and the abundance becomes larger. In the red shaded region of Fig.~\ref{fig:m2case12}$\,$({\bf right}), $N_3$ decays after it dominates the universe for the maximum value of $M_1 = M_3$; smaller $M_1$ enlarges the region. In the union of orange and red shaded regions, $N_3$ creates entropy after BBN, which is excluded since the baryon abundance at BBN and at CMB would differ. We see that no parameter region is then consistent with $|y_{23}|^2 v^2 / M_3$ being larger than the active neutrino mass, completing the proof.

Since $|y_{23}|^2 v^2 / M_3$ is at the most as large as the active neutrino mass, with the upper bound on $|y_{13}|$ from the stability of $N_1$, $m_{23}$ is much smaller than the active neutrino mass. The active neutrino mass is almost 2 by 2, showing {\bf Claim 1}.

The range of $c M_2 (v/v_R)^2$ is constrained. It cannot be much larger than the active neutrino mass; since  $|y_{23}|^2 v^2 / M_3$ is at the most as large as the active neutrino mass, $m_{22}$ cannot be fine-tuned to be small enough. If $c M_2 (v/v_R)^2$ is much smaller than the active neutrino mass, $c M_3 (v/v_R)^2$ is also small. Then the active neutrino mass matrix is dominantly given by the see-saw from $M_3$, and two active neutrinos remain massless, which is in contradiction with observations.  The only possibility is that $c M_2 (v/v_R)^2$ is comparable to the active neutrino mass, showing {\bf Claim 2}.

\vspace{0.5cm}

We next consider the case with $M_3 < M_1$. The term $c (v/v_R)^2 M_3$ is much smaller than the observed neutrino masses. The active neutrino mass matrix is dominantly given by the term $c (v/v_R)^2 M_2$ and the seesaw from $N_3$, or a Dirac mass term with $N_3$ if $M_3 \ll 0.1$ eV, and hence is essentially rank-2. The lightest active neutrino mass is much lighter than $0.01$ eV. If $c (v/v_R)^2 M_2$ is smaller than the observed active neutrino mass, the active neutrino mass matrix is essentially rank-1 and cannot explain the observed active neutrino mass. Thus, it is required that $c (v/v_R)^2 M_2 \gsim \mu$. It is possible that $c (v/v_R)^2 M_2 \gg \mu$ if it is cancelled by $|y_{23}|^2 v^2 / M_3$.
The constraint on the case with $c (v/v_R)^2 M_2 \gg \mu$ is obtained by interpreting Fig.~\ref{fig:thermal_FO} with $c$ smaller than the actual value of $c$.

\section{A symmetry for the cosmological stability of $N_1$}
\label{sec:N1stab}

For $N_1$ to make up dark matter, the mixing of active and sterile neutrinos must be very small to avoid limits on the radiative decay $N_1 \rightarrow \nu \gamma$, as shown in (\ref{eq:sin2theta}). Sufficient stability can be a natural if a symmetry forbids the $\ell N_1 H_L$ interaction in the effective theory (\ref{eq:LREFT}), so that $y_{1i} = 0$.  Any LR theory giving an effective theory below $v_R$ with no interactions for $N_i$ is particularly interesting: not only is $N_1$ cosmologically stable, but if $N_2$ has a mass significantly less than $v_R$, it is necessarily long-lived with a lifetime governed by $W_R$-mediated beta decay.  In this case the allowed values of $v_R$ and $M_1$ are correlated - it is necessary to be on the blue line of Fig.~\ref{fig:thermal_FO} rather than in the unshaded triangle.  Furthermore, since $N_2$ has a 10\% branching ratio to decay to $N_1$, there is a component of DM that is hot, becoming non-relativistic around the eV era, with  $\Delta N_{\rm eff} \sim 0.1$ and $m_{\nu,{\rm eff}} \sim 1.1 \,\EV$. As described in Sec.~\ref{sec:hotness}, and shown in Fig.~\ref{fig:signals}, this is close to present limits and will be discovered or refuted by CMB Stage IV~\cite{Abazajian:2016yjj}.

For a LR model based on Higgs doublets $H_{L,R}$, such a symmetry must forbid the operator $\ell \bar{\ell} \, H_L H_R$, which leads to $\ell N H_L$, while allowing $\ell  \bar{\ell} \, H^\dagger_L H^\dagger_R$, which yields the charged lepton Yukawa couplings $\ell \bar{e} H^\dagger_L$, as well as the Majorana mass operators $\ell \ell \, H_L H_L$ and $\bar{\ell}  \bar{\ell} \, H_R H_R$.  For example, this could be accomplished by a $Z_{4L} \times Z_{4R}$ symmetry with $\ell$ and $H_L$ transforming as  $(i, 1)$ and  $(\bar{\ell}, H_R)$ as $(1, i)$.  The operator $q \bar{q} H_L H_R$ or $q \bar{q} H_L^\dag H_R^\dag$ is inconsistent with this $Z_{4L} \times Z_{4R}$ symmetry, so that the down and/or up-type quark Yukawa couplings must be generated by a different set of doublets, $H^{(q)}_{L,R}$, with the effective theory below $v_R$ containing the two doublets $H_L$ and $H^{(q)}_L$. A weak-scale Nambu-Goldstone boson is avoided by introducing a soft breaking of the $Z_{4L} \times Z_{4R}$ symmetry via the mass operator  $H^\dagger_LH^{(q)}_L$.

\bibliographystyle{JHEP}
\bibliography{LRbib}

\end{document}